\documentclass[twocolumn,twocolappendix]{aastex63}

\def\mgii{\ion{Mg}{2}}
\def\civ{\ion{C}{4}}
\def\cii{\ion{C}{2}}
\def\nv{\ion{N}{5}}
\def\heii{\ion{He}{2}}

\def\feii{\ion{Fe}{2}}

\hypersetup{linkcolor=red,citecolor=blue,filecolor=cyan,urlcolor=magenta}

\usepackage{braket}
\usepackage{amsmath}
\usepackage{bm}
\received{}
\revised{}
\accepted{}

\shorttitle{NIR Spectroscopy of a Large $z \gtrsim 6.5$ Quasar Sample}
\shortauthors{Yang et al.}

\begin{document}

\title{Probing Early Super-massive Black Hole Growth and Quasar Evolution with Near-infrared Spectroscopy of 37 Reionization-era Quasars at $6.3 < z \le 7.64$}

\correspondingauthor{Jinyi Yang}
\email{jinyiyang@email.arizona.edu}

\author[0000-0001-5287-4242]{Jinyi Yang}
\altaffiliation{Strittmatter Fellow}
\affil{Steward Observatory, University of Arizona, 933 N Cherry Ave, Tucson, AZ 85721, USA}

\author[0000-0002-7633-431X]{Feige Wang}
\altaffiliation{NHFP Hubble Fellow}
\affil{Steward Observatory, University of Arizona, 933 N Cherry Ave, Tucson, AZ 85721, USA}

\author[0000-0003-3310-0131]{Xiaohui Fan}
\affil{Steward Observatory, University of Arizona, 933 N Cherry Ave, Tucson, AZ 85721, USA}

\author[0000-0002-3026-0562]{Aaron J. Barth}
\affil{Department of Physics and Astronomy, University of California, Irvine, CA 92697, USA}

\author[0000-0002-7054-4332]{Joseph F. Hennawi}
\affil{Department of Physics, University of California, Santa Barbara, CA 93106-9530, USA}

\author[0000-0002-2579-4789]{Riccardo Nanni}
\affil{Department of Physics, University of California, Santa Barbara, CA 93106-9530, USA}

\author[0000-0002-1620-0897]{Fuyan Bian}
\affil{European Southern Observatory, Alonso de C\'ordova 3107, Casilla 19001, Vitacura, Santiago 19, Chile}

\author[0000-0003-0821-3644]{Frederick B. Davies}
\affil{Lawrence Berkeley National Laboratory, 1 Cyclotron Rd, Berkeley, CA 94720-8139, USA}

\author[0000-0002-6822-2254]{Emanuele P. Farina}
\affil{Max Planck Institut f\"ur Astrophysik, Karl--Schwarzschild--Stra{\ss}e 1, D-85748, Garching bei M\"unchen, Germany}

\author[0000-0002-4544-8242]{Jan-Torge Schindler}
\affil{Max Planck Institut f\"ur Astronomie, K\"onigstuhl 17, D-69117, Heidelberg, Germany}

\author[0000-0002-2931-7824]{Eduardo Ba\~nados}
\affil{Max Planck Institut f\"ur Astronomie, K\"onigstuhl 17, D-69117, Heidelberg, Germany}

\author[0000-0002-2662-8803]{Roberto Decarli}
\affil{INAF -- Osservatorio di Astrofisica e Scienza dello Spazio di Bologna, via Gobetti 93/3, I-40129 Bologna, Italy}

\author[0000-0003-2895-6218]{Anna-Christina Eilers}
\altaffiliation{NHFP Hubble Fellow}
\affil{MIT Kavli Institute for Astrophysics and Space Research, 77 Massachusetts Ave., Cambridge, MA 02139}

\author[0000-0003-1245-5232]{Richard Green}
\affil{Steward Observatory, University of Arizona, 933 N Cherry Ave, Tucson, AZ 85721, USA}

\author[0000-0001-8416-7059]{Hengxiao Guo}
\affil{Department of Physics and Astronomy, University of California, Irvine, CA 92697, USA}

\author[0000-0003-4176-6486]{Linhua Jiang}
\affil{Kavli Institute for Astronomy and Astrophysics, Peking University, Beijing 100871, China}

\author[0000-0001-6239-3821]{Jiang-Tao Li}
\affil{Department of Astronomy, University of Michigan, 311 West Hall, 1085 S. University Ave, Ann Arbor, MI, 48109-1107, USA}

\author[0000-0001-9024-8322]{Bram Venemans}
\affil{Max Planck Institut f\"ur Astronomie, K\"onigstuhl 17, D-69117, Heidelberg, Germany}

\author[0000-0003-4793-7880]{Fabian Walter}
\affil{Max Planck Institut f\"ur Astronomie, K\"onigstuhl 17, D-69117, Heidelberg, Germany}

\author[0000-0002-7350-6913]{Xue-Bing Wu}
\affil{Kavli Institute for Astronomy and Astrophysics, Peking University, Beijing 100871, China}
\affil{Department of Astronomy, School of Physics, Peking University, Beijing 100871, China}

\author[0000-0002-5367-8021]{Minghao Yue}
\affil{Steward Observatory, University of Arizona, 933 N Cherry Ave, Tucson, AZ 85721, USA}



\begin{abstract}
We report the results of near-infrared spectroscopic observations of 37 quasars in the redshift range $6.3< z\le7.64$, including 32 quasars at $z>6.5$, forming the largest quasar near-infrared spectral sample at this redshift. The spectra, taken with Keck, Gemini, VLT, and Magellan, allow investigations of central black hole mass and quasar rest-frame ultraviolet spectral properties. The black hole masses derived from the \mgii\ emission lines are in the range $(0.3-3.6)\times10^{9}\,M_{\odot}$, which requires massive seed black holes with masses $\gtrsim10^{3-4}\,M_{\odot}$, assuming Eddington accretion since $z=30$. The Eddington ratio distribution peaks at $\lambda_{\rm Edd}\sim0.8$ and has a mean of 1.08, suggesting high accretion rates for these quasars. The \civ\ -- \mgii\ emission line velocity differences in our sample show an increase of \civ\ blueshift towards higher redshift, but the evolutionary trend observed from this sample is weaker than the previous results from smaller samples at similar redshift. The \feii/\mgii\ flux ratios derived for these quasars up to $z=7.6$, compared with previous measurements at different redshifts, do not show any evidence of strong redshift evolution, suggesting metal-enriched environments in these quasars. Using this quasar sample, we create a quasar composite spectrum for $z>6.5$ quasars and find no significant redshift evolution of quasar broad emission lines and continuum slope, except for a blueshift of the \civ\ line. Our sample yields a strong broad absorption line quasar fraction of $\sim$24\%, higher than the fractions in lower redshift quasar samples, although this could be affected by small sample statistics and selection effects.
\end{abstract}

\keywords{quasars: general - quasars: emission lines - quasars: supermassive black holes}


\section{Introduction} \label{sec:intro}
Observations of high-redshift ($z > 6$) quasars hold the key to understanding the formation and evolution of the earliest supermassive black holes (SMBHs) and galaxies. Recent observations of quasars at $z>6$ have revealed the existence of massive SMBHs with $\sim 10^{8} - 10^{10}$ solar masses in a very young Universe \citep[e.g.,][]{wu15,banados18,matsuoka19b,onoue19,shen19,yang20a,wang21b}, within only 920 million years of the Big Bang. This raises the question of how these SMBHs grow to a few billion solar mass within such short time. Theoretical models with different seed black hole (BH) mass and/or different modes of accretion offer several potential explanations of the formation and growth of early SMBHs.
Detailed observations of a large sample of the highest redshift quasars are needed to test these models and to improve our understanding of SMBH formation and evolution. 
Such studies rely on both  wide-field high-redshift quasar surveys for the discovery and high-quality spectroscopic observations of high-redshift quasars at optical and near-infrared (NIR) wavelengths to measure quasar properties. 

Recent progress in deep imaging surveys coupled with NIR spectroscopic capabilities on large telescopes have significantly increased the sample size of $z>6$ quasars to $\sim$ 200 and pushed the quasar redshift frontier to $z\gtrsim 7.5$, deep into the epoch of reionization \citep[e.g.,][]{mortlock11,jiang16,mazzucchelli17,banados18,fan19,reed19,matsuoka19a,matsuoka19b,venemans13,venemans15,wang18, wang19,yang19b,yang20a,wang21b}. \citet[][hereafter W19]{wang19} and \citet[][hereafter Y19]{yang19b} have recently carried out a new wide-field survey for reionization-era quasars in a $\sim$20,000 deg$^2$ area by combining a number of publicly available deep  optical and infrared photometric datasets. This survey has already discovered more than 35 quasars at $6.3 < z \le 7.64$ \citep[also,][]{fan19,wang18, yang20a, wang21b}. These successful surveys of high-redshift quasars have significantly expanded the high-redshift quasar sample and provided valuable new targets for the investigations of both reionization history \cite[e.g.,][]{davies18,yang20a,yang20b,wang20} and early SMBHs. 

The measurements of BH masses in high-redshift ($z > 6$) quasars are mainly based on the quasar \mgii\ emission line from NIR spectra, since \mgii\ is the best tracer in the observable wavelength range (i.e., optical and NIR). Combined with the bolometric luminosity derived from the NIR spectra after applying bolometric corrections, the measurement of BH mass allows us to estimate the Eddington ratio of these SMBHs. By fitting the continuum of NIR spectra and the \mgii\ emission line, the BH mass and Eddington ratio of a number of $z > 6$ quasars have been derived \citep[e.g.,][]{jiang07, kurk07, willott10, derosa14, mazzucchelli17, onoue19, shen19, schindler20}. These measurements have improved our understanding of BH growth and accretion in the early Universe and also raised questions related to early SMBH formation, accretion, and BH-host galaxy co-evolution. At the same time, NIR spectroscopy also allows studies of the rest-frame ultraviolet (UV) properties of these early quasars. The evolution of quasar spectral properties (e.g., broad emission line velocity shifts) gives insight into the physical conditions and emission mechanisms of the quasar broad-line region (BLR) \citep[e.g.,][]{gaskell82, richards11,meyer19}. In particular, the \feii/\mgii\ ratio traces the chemical abundances in the quasar BLR and is an important diagnostic of the iron enrichment and the history of star formation in quasar host galaxies in the early Universe \citep[e.g.,][]{hamann99, jiang07, derosa11, schindler20, onoue20}. 

We conducted a NIR spectroscopic survey of quasars selected from a new survey (W19 and Y19) and other known $z > 6.5$ quasars that did not have published NIR spectra before our observations. In this paper, we present the NIR spectral dataset, including spectra of 37 quasars at $6.3 < z \le 7.64$, and the results obtained from its analysis. 
We describe the NIR spectral dataset including the quasar sample, observations, and data reduction in Section 2. The spectral analysis is presented in Section 3. We report the measurements of BH mass and Eddington ratio in Section 4 and discuss the quasar rest frame UV spectral properties in Section 5. We then discuss early SMBH growth and broad absorption line (BAL) quasars in this sample in Section 6. A summary of this work is presented in Section 7.
All results below refer to a $\Lambda$CDM cosmology with parameters $\Omega_{\Lambda}$ = 0.7, $\Omega_{m}$ = 0.3, and $h$ = 0.7. 
 
\section{The NIR Dataset}
\subsection{Quasar Sample}
Our NIR spectroscopic observations mainly target the new quasars from a series of recent investigations \citep{fan19, wang18,wang19,wang21b,yang19b, yang20a} and also include some previously known $z>6.5$ quasars that did not have published NIR spectra before our observations. The NIR spectral sample presented in this paper is constructed based on (1) quasars from the survey described in W19 and Y19, (2) other known $z >7$ quasars (i.e., J1120+0641 and J1342+0928), (3) quasars observed in our Keck/NIRES NIR spectroscopic programs, and (4) other $z>6.5$ quasars (i.e., J0024+3913 and J2232+2930) that are not in the first three categories but have Gemini/GNIRS data in the archive. The final sample includes 37 quasars at $6.3 < z \le 7.64$. 
Within this sample, ten quasars have been published with BH mass measurements in the literature \citep{mortlock11,venemans15, banados18,wang18,fan19,tang19,wang20,yang20a,banados21,wang21b}.
We include them and present our new BH mass measurements of these quasars in this paper to compare all quasar properties consistently. 

In this sample, there are six previously unpublished quasars. They are newly discovered objects found in an ongoing survey, based on the same selection method as used previously in W19 and Y19. W19 conducted a $z \gtrsim 6.5$ quasar survey based on color-color selection using photometric data from the DESI Legacy imaging Surveys \citep[DELS,][]{dey18}, Pan-STARRS1 \citep[PS1,][]{chambers16}, and all public NIR imaging surveys, as well as the Wide-field Infrared Survey Explorer \citep[{\it WISE},][]{wright10} mid-infrared survey in the northern sky, while Y19 carried out a similar quasar survey in the southern sky using the data from the Dark Energy Survey \citep[DES,][]{abbott18} DR1 instead of DELS and PS1. 
In this paper, we report the coordinates and the NIR spectra of these six new quasars. They are J021847.039+000715.175, J052559.675--240622.98, J092358.997+075349.107, J105807.720+293041.703, J200241.594--301321.69, and J233807.032+214358.17. The quasar selection, discovery, and other properties will be presented in detail in a separate paper. Table \ref{tab:sample} lists the full sample of 37 quasars and their redshifts.

\begin{figure}
\centering 
\epsscale{1.15}
\plotone{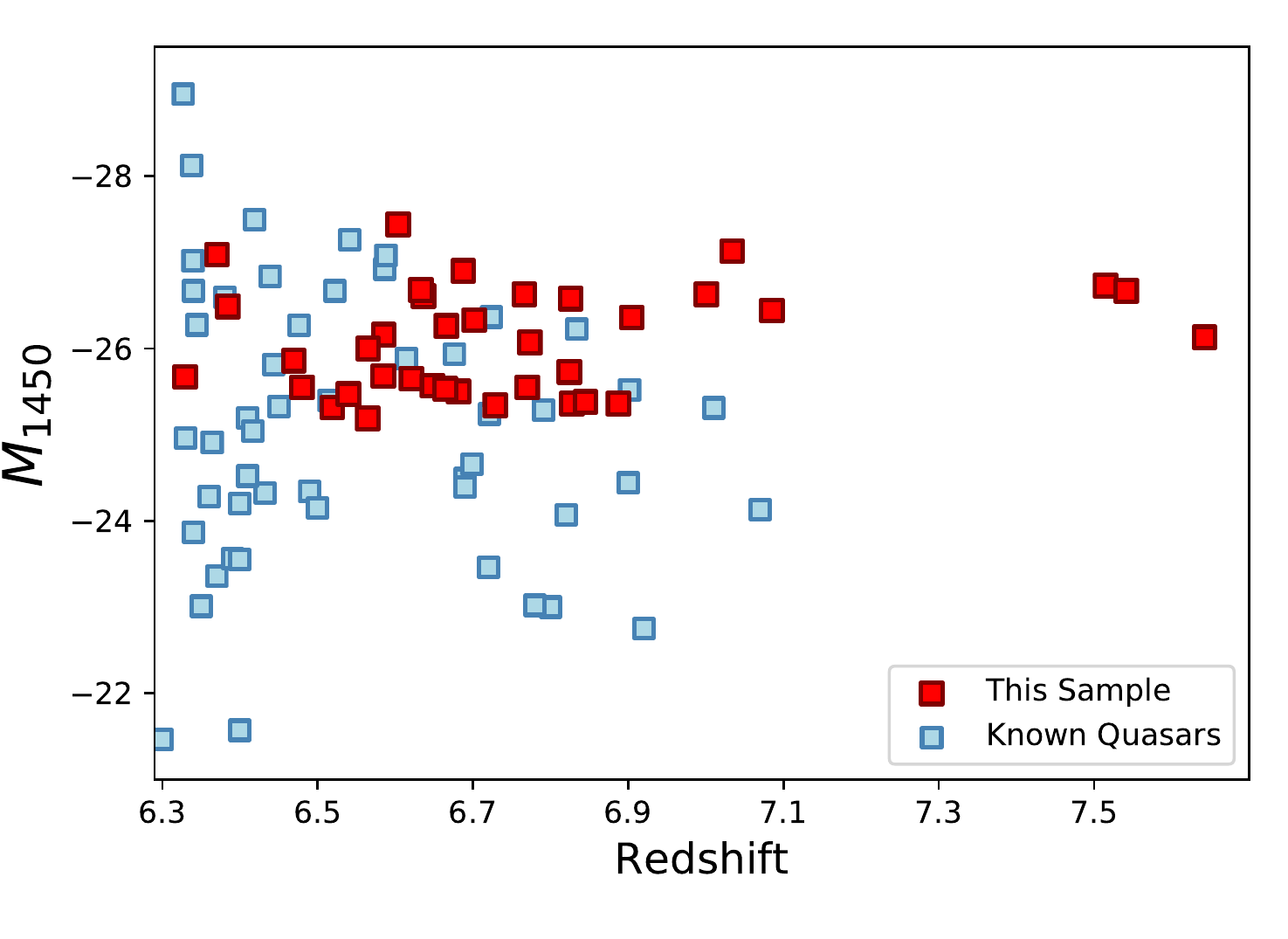} 
\caption{The redshift and absolute magnitude distribution of quasars in our sample (red squares) and other known quasars at $z>6.3$ (blue squares). This new NIR spectroscopic sample covers 37 quasars from redshift 6.35 to the most distant known one at $z=7.64$, which forms the largest NIR spectral dataset for quasars at $z>6.5$.}
\label{fig:sample}
\end{figure}

As shown in Figure \ref{fig:sample}, our new NIR spectral sample comprises the largest NIR spectral dataset of quasars at $z>6.5$ (32 quasars). Thus the derived measurements from these quasars will be representative of the observed NIR properties of luminous quasars in this redshift range, in an absolute magnitude range $M_{\rm 1450} < -25.2$.
Moreover, this NIR sample contains a subsample of 32 quasars that meet the uniform selection used in W19 and Y19. These quasars therefore form a complete sample for the measurement of the BH mass function at $z\sim 6.5$ (J. Yang et al. in prep).

\begin{deluxetable*}{ l l l l l l l l l l}
\tablecaption{Quasar Information and Observation Information of the 37 Quasars in Our Sample.}
\tabletypesize{\scriptsize}
\tablewidth{0pt}
\tablehead{
\colhead{Name} &
\colhead{Instrument} &
\colhead{Exp Time (s)} &
\colhead{$z$} &
\colhead{$z_{\rm err}$} &
\colhead{Discovery} &
\colhead{[\cii]} &
\colhead{$z$\_Ref\tablenotemark{a}} &
\colhead{NIR\tablenotemark{b}} &
\colhead{$J$ (AB)\tablenotemark{c}}
}
\startdata
  J002429.77+391319.0\tablenotemark{d} & GNIRS & 13800 & 6.621 & 0.002 & \cite{tang17} & Y & \cite{mazzucchelli17} & Y & 20.77$\pm$0.15\\ 
  J003836.10$-$152723.6 & GNIRS & 15300 & 7.0340 & 0.0003 & \cite{wang18} & Y & Wang in prep & Y & 19.69$\pm$0.07\\ 
  J004533.57+090156.9\tablenotemark{d} & NIRES & 13680 & 6.4694 & 0.0025 & \cite{mazzucchelli17} & Y & \cite{eilers20} & N & 20.80$\pm$0.13\\ 
  J021847.04+000715.2\tablenotemark{e} & NIRES & 5760 & 6.7700 & 0.0013 & Yang in prep & Y & Wang in prep & N & 21.08$\pm$0.30\\ 
  J024655.90$-$521949.9 & X-Shooter & 24000& 6.8876 & 0.0003 & Y19 & Y & Wang in prep & N & 21.29$\pm$0.19\\ 
  J025216.64$-$050331.8& NIRES/X-Shooter & 18000/28800 & 7.0006 & 0.0009 & Y19 & Y & Wang in prep & Y & 20.19$\pm$0.07\\ 
  J031343.84$-$180636.4  & FIRE/F2/ & 21723/11040/ & 7.6423 & 0.0013 & \cite{wang21b}& Y & Wang in prep & Y & 20.94$\pm$0.13\\ 
                           &GNIRS/NIRES & 29100/16200 & & & & & & & \\
  J031941.66$-$100846.0 & NIRES & 18720 & 6.8275 & 0.0021 & Y19 & Y & Wang in prep & N & 20.98$\pm$0.24\\ 
  J041128.63$-$090749.8 & NIRES & 5760 & 6.8260 & 0.0007 & W19 & Y & Wang in prep & N & 20.02$\pm$0.14\\ 
  J043947.08+163415.7 & GNIRS & 3600 & 6.5188 & 0.0004 & \cite{fan19} & Y & \cite{yang19a} & Y & 17.46$\pm$0.02\\ 
  J052559.68$-$240623.0\tablenotemark{e}  & F2 & 5400 & 6.5397 & 0.0001 & Yang in prep & Y & Wang in prep & N & ---\\ 
  J070626.39+292105.5 & NIRES & 15210 & 6.6037 & 0.0003 & W19 & Y & Wang in prep & N & 19.16$\pm$0.05\\ 
  J080305.42+313834.2 & GNIRS & 3600 & 6.377 & 0.006 & W19 & N & W19 & N & 20.12$\pm$0.12\\ 
  J082931.97+411740.4 & GNIRS & 13500 & 6.768 & 0.006 & W19 & N & W19 & N & 20.28$\pm$0.15\\ 
  J083737.84+492900.4 & GNIRS & 17400 & 6.710 & 0.008 & W19 & N & W19 & N & 20.21$\pm$0.17\\
  J083946.88+390011.5 & GNIRS & 16800 & 6.905 & 0.01 & W19 & N & W19 & N & 20.39$\pm$0.20\\ 
  J091054.53$-$041406.8 & GNIRS/NIRES & 3600/3600 & 6.6363 & 0.0003 & W19 & Y & Wang in prep & N & 20.25$\pm$0.14\\ 
  J091013.63+165629.8 & GNIRS & 13200 & 6.7289 & 0.0005 & W19 & Y & Wang in prep & N & 21.06$\pm$0.13\\ 
  J092120.56+000722.9 & GNIRS & 9600 & 6.5646 & 0.0003 & \cite{matsuoka18} & Y & Wang in prep & N & 21.21$\pm$0.28\\ 
  J092347.12+040254.4 & NIRES & 11880 & 6.6330 & 0.0003 & W19,\cite{matsuoka18} & Y & Wang in prep & N & 20.02$\pm$0.09\\ 
  J092359.00+075349.1\tablenotemark{e}  & GNIRS & 7200 & 6.6817 & 0.0005 & Yang in prep & Y & Wang in prep & N & ---\\ 
  J100758.26+211529.2 & GNIRS/NIRES & 21900/7920 & 7.5149 & 0.0004 & \cite{yang20a} & Y & \cite{yang20a} & Y & 20.22$\pm$0.18\\ 
  J105807.72+293041.7\tablenotemark{e}  & NIRES & 3600 & 6.5846 & 0.0005 & Yang in prep & Y & Wang in prep & N & ---\\ 
  J110421.59+213428.8 & GNIRS & 7200 & 6.7662 & 0.0009 & W19 & Y & Wang in prep & N & 19.95$\pm$0.12\\ 
  J112001.48+064124.3 & GNIRS & 4800 & 7.0851 & 0.0005 & \cite{mortlock11} & Y & \cite{venemans17a} & Y & 20.35$\pm$0.15\\ 
  J112925.34+184624.2\tablenotemark{d} & NIRES & 12600 & 6.823 & 0.003 & \cite{banados21} & N & \cite{banados21} & Y & 20.90$\pm$0.11\\ 
  J113508.93+501133.0 & GNIRS/NIRES & 7200/4800 & 6.5851 & 0.0008 & W19 & Y & Wang in prep & N & 20.41$\pm$0.16\\ 
  J121627.58+451910.7 & GNIRS & 4800 & 6.65 & 0.01 & W19 & N & W19 & N & 21.02$\pm$0.13\\ 
  J131608.14+102832.8 & NIRES & 3000 & 6.35 & 0.04 & W19 & N & W19 & N & 20.75$\pm$0.12\\ 
  J134208.10+092838.6 & GNIRS & 32400 & 7.5413 & 0.0007 & \cite{banados18} & Y & \cite{venemans17b} & Y & 20.30$\pm$0.02\\ 
  J153532.87+194320.1 & NIRES & 2880 & 6.40 & 0.05 & W19 & N & W19 & N & 19.64$\pm$0.11\\ 
  J172408.74+190143.0\tablenotemark{d} & NIRES & 15120 & 6.44 & 0.05 & \cite{mazzucchelli17} & N & \cite{mazzucchelli17} & N & 21.09$\pm$0.18\\ 
  J200241.59$-$301321.7\tablenotemark{e}  & GNIRS & 3600 & 6.6876 & 0.0004 & Yang in prep & Y & Wang in prep & N & 19.97$\pm$0.16\\ 
  J210219.22$-$145854.0 & GNIRS/NIRES & 10200/5760 & 6.6645 & 0.0002 & W19 & Y & Wang in prep & N & 21.14$\pm$0.20\\ 
  J221100.60$-$632055.8 & X-Shooter & 31200 & 6.8449 & 0.0003 & Y19 & Y & Wang in prep & N & 21.23$\pm$0.18\\ 
  J223255.15+293032.0\tablenotemark{d} & GNIRS & 4800 & 6.666 & 0.004 & \cite{venemans15} & Y & \cite{mazzucchelli17} & Y & 20.37$\pm$0.14\\ 
  J233807.03+214358.2\tablenotemark{e}  & GNIRS/NIRES & 1500/7200 & 6.60 & 0.03 & Yang in prep & N & Yang in prep & N & 20.75$\pm$0.30\\ 
 \enddata
 \tablenotetext{a}{The reference for the redshifts used in the spectral analysis. If the quasar has a [\cii] detection (column [\cii] = Y), the reference is for the [\cii]-based redshift, and the redshift listed in column $z$ is the [\cii]-based redshift. Most of the [\cii] detections are from a series of  ALMA/NOEMA programs that will be reported in detail in F. Wang et al. (in prep).}
  \tablenotetext{b}{The NIR column reports whether the object has previously published BH mass measurements (Y or N).}
   \tablenotetext{c}{The $J$-band photometric data used to scale the NIR spectra. For the quasars J0525--2406, J0923+0753, and J1058+2930 without $J$ data, we used $Y$ or $K$-band photometry, as described in Section 2.3.}
    \tablenotetext{d}{These quasars are also named as PSO J006.1240+39.2219, PSO J011.3898+09.0324, PSO J172.3556+18.7734, PSO J261.0364+19.0286, and PSO J338.2298+29.5089, respectively.}
 \tablenotetext{e}{These quasars are previously unpublished.  Details of their selection and identification will be reported separately (J. Yang et al. in prep).}
 \label{tab:sample}
\end{deluxetable*}

\subsection{NIR Spectroscopy}
We obtained NIR spectroscopy of our quasar sample using the following facilities: Keck/NIRES (Near-Infrared Echellette Spectrometer, \citealt{elias06a,elias06b}), Gemini/GNIRS (Gemini Near-Infrared Spectrograph, \citealt{wilson04}), VLT/X-Shooter \citep{vernet11}, Gemini/F2 (FLAMINGOS-2 near-infrared imaging spectrograph, \citealt{eikenberry04}), and Magellan/FIRE (Folder-port InfraRed Echellette, \citealt{simcoe10}). Table \ref{tab:sample} lists the instruments used to observe each quasar and the exposure times, and the observations with each instrument are described below.
\begin{itemize}
\vspace{-8pt}
\setlength{\itemsep}{-0.5em}
\item[1)] We observed 18 quasars with Keck/NIRES from 2018 to 2020. Keck/NIRES has a fixed configuration that simultaneously covers 0.94 to 2.45 $\mu$m with a fixed $0\farcs55$ narrow slit, resulting in a resolving power of $R\sim2700$.
\item[2)] Spectra of 22 quasars were taken with Gemini North/GNIRS, including 18 quasars observed in our programs from 2018 to 2020 and four from Gemini archival data.
We used the short-slit (cross-dispersion) mode (32 l/mm) with simultaneous coverage of 0.85--2.5 $\mu$m. A 0$\farcs$675 slit was used, corresponding to $R\sim700$. 
For the archival data, three of the quasars were observed with a 0$\farcs$675 slit, and one (J2232+2930) was observed using a 1$\farcs$0 slit ($R\sim500$).
\item[3)] In addition, we observed three quasars with VLT/X-Shooter (ID: 0103.A-0423(A)) in 2019. X-Shooter covers the wavelength range from 3000 to 24800 \AA. We used a 0$\farcs$9 slit for the VIS (5595-10240 \AA) and a 0$\farcs$6 slit for NIR (10240-24800 \AA), resulting in resolving power of 8900 and 8100, respectively. 
\item[4)] Quasars J0313--1806 and J0525--2406 were observed with Gemini South F2 in 2019. For both, we used a slit width of 0$\farcs$72 which delivers a spectral resolving power of $R\sim400$. With F2, only an $HK$ range spectrum was obtained, covering the wavelength range from 1.45 to 2.5 $\mu$m.
\item[5)] In addition to the NIRES, GNIRS, and F2 observations, quasar J0313--1806 was also observed with Magellan/FIRE (0.8-2.5 $\mu$m) in Echelle mode in 2019 November and December with 0$\farcs$75 and 1$\farcs$0 slits, corresponding to resolving power of $R\sim4800$ and $R\sim3600$, respectively. 
There are seven quasars observed with multiple instruments.
\end{itemize}
 
\subsection{Data Reduction}
All NIR spectra are reduced with the open-source Python-based spectroscopic data reduction pipeline {\tt PypeIt}\footnote{\url{https://github.com/pypeit/PypeIt}} \citep{prochaska20}. 
The wavelength solutions are derived from the night sky OH lines in the vacuum frame. We choose this method in order to use on-sky wavelength calibrations and also to reduce observational overheads given that our science goals do not require very high resolution.
The sky subtraction is based on the standard A--B mode and a b-spline fitting procedure that is performed to further clean up the sky line residuals following \cite{bochanski09}. An optimal extraction \citep{horne86} is performed to generate 1D science spectra. The extracted spectra are flux calibrated with sensitivity functions derived from the observations of spectroscopic standard stars.
Telluric absorption is corrected by fitting absorption models to the quasar spectra, and the absorption models are constructed using a telluric model plus a quasar model.
The telluric model grids are produced from the Line-By-Line Radiative Transfer Model \citep[{\tt LBLRTM}\footnote{\url{http://rtweb.aer.com/lblrtm.html}};][]{clough05}.
The quasar model is based on a principal component analysis method \citep{davies18}. 

The stacking of individual exposures or spectra from multiple instruments does not employ any interpolation to avoid correlated noise. 
We determine a common wavelength grid based on the dispersion of each instrument. The wavelength grid is sampled linearly in velocity space for echelle spectrographs and linearly in wavelength for other long-slit spectrographs. For spectra from multiple instruments, the wavelength grid is derived based on the lowest resolution spectrum. 
We then use a histogram technique to divide all native pixels into wavelength bins. The stacked flux in each wavelength bin is then computed as the mean flux density of values from all native pixels in that bin, weighted by the average square of the signal-to-noise ratio (S/N) of the exposure that contains this pixel. 

The reduced spectrum of each quasar is then scaled using its $J$-band magnitude (or scaled with $K$ or $Y$ if $J$ band is not available). We choose $J$ band instead of the $K$ band, which includes the quasar \mgii\ emission line, since only a few objects have $K$-band photometric data. For the quasars (23 objects) included in public $J$-band photometric catalogs from the UKIRT Hemisphere Survey \citep[UHS;][]{dye18}, the UKIRT InfraRed Deep Sky Surveys--Large Area Survey \citep[ULAS;][]{lawrence07}, or the VISTA Hemisphere Survey \citep[VHS;][]{mcmahon13}, we use the public $J$-band data. For quasars not in these catalogs but which have published $J$-band magnitudes (8 objects), we use the corresponding $J$-band data. Then for the quasars without either of those sources of $J$-band data, if they have been observed in our UKIRT/WFCAM imaging programs \citep{wang19}, we use photometric data from our UKIRT images (2 objects). For those quasars that do not have any of the $J$ data described above but are covered by $J$-band images from public surveys, we perform forced photometry (3\arcsec\ diameter) on the public $J$-band images. 

\begin{figure*}
\centering 
\epsscale{1.18}
\plotone{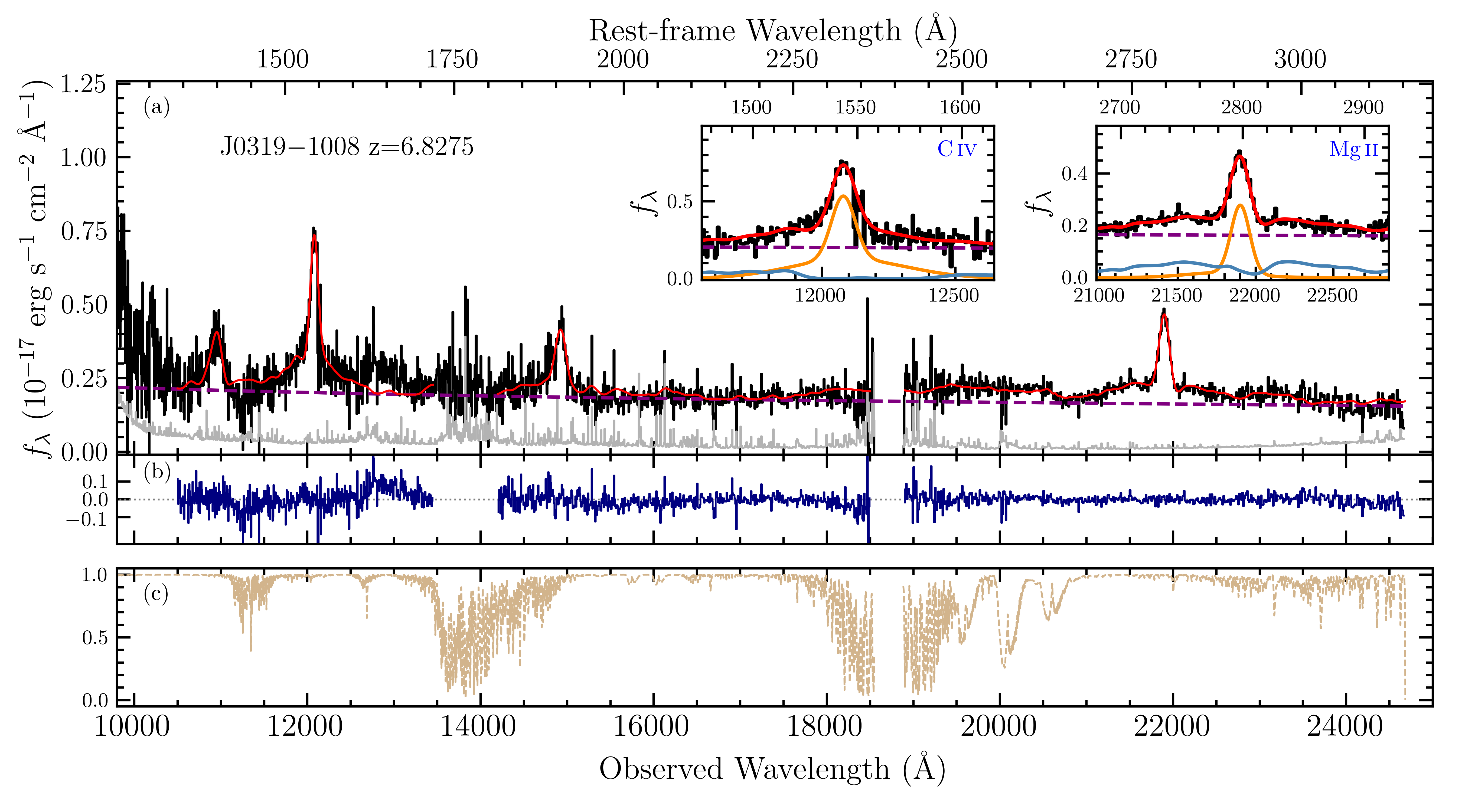} 
\caption{An example of spectral fitting for the quasar J0319--1008 at $z = 6.8$. {\bf (a)} The spectrum (black line) is acquired with Keck/NIRES. The grey line shows the spectral uncertainty. The purple dashed line represents the best-fit power law continuum and the solid red line denotes the total fit. The two inset plots show the fits to the \civ\ and \mgii\ emission lines. The orange and blue solid lines represent the best-fit emission line and iron components, respectively. The quasar J0319--1008 has a [\cii]-based redshift of 6.8275$\pm$0.0021 (F. Wang et al. in prep). The line fitting yields a \mgii-based redshift of 6.816$\pm$0.004. The continuum fitting obtained a power-law slope of $\alpha_{\rm \lambda}=-0.45\pm0.3$. The fitting of all other quasar spectra is shown in Figure \ref{fig:allspec01} in Appendix A. {\bf (b)} The residual (data -- model) of spectral fitting. {\bf (c)} The telluric model used for the telluric correction for this quasar.}
\label{fig:fitting}
\end{figure*}

The quasars J0923+0753 and J1058+2930 have only $< 3\sigma$ forced photometric magnitudes in $J$ band. For J0923+0753, we use its forced photometric data in the $Y$ band (21.25$\pm$0.26 in AB), which has S/N $>4\sigma$. For the quasar J1058+2930, which does not have images in any NIR bands, we use the photometry from the acquisition image from NIRES in the $K_s$ band (20.56$\pm$0.05 in AB). As a comparison, we also use the same scaling factor as that for the quasar J0910--0414, which was observed with NIRES just a few hours before J1058+2930 on the same night. Both yield consistent scaling factors, with a 5\% difference in flux density.
The quasar J0525--2406 has only an F2 spectrum covering the $H$ and $K$ bands, and it does not have public NIR images. For this object we take $K_s$-band imaging with MMT/MMIRS and scale its spectrum using this $K$-band photometry (20.48$\pm$0.09 in AB). Then, all scaled spectra are corrected for Galactic extinction based on the dust map of \cite{schlegel98} and the extinction law of \cite{cardelli89}.
All photometric data used to scale spectra are listed in Table \ref{tab:sample}.  All final spectra are shown in Figure \ref{fig:fitting} and Figure \ref{fig:allspec01}. Since this work focuses on the NIR range, we plot the spectra only at wavelengths redder than 9800 \AA. 

\section{Spectral Analysis}
After obtaining the final spectral dataset, we fit each quasar spectrum to derive spectral properties for further measurements. In this section, we describe our spectral fitting procedure for the quasar continuum and emission lines. We will also discuss a few individual quasars with unusual spectral features.

\subsection{Spectral Fitting}
We fit each near-IR spectrum with a model consisting of continuum plus emission lines. The initial redshift is chosen to be the [\cii]-based redshift or the published redshift as listed in Table \ref{tab:sample}. The pseudo-continuum includes a power-law continuum, \feii\ template \citep{vestergaard01,tsuzuki06}, and Balmer continuum \citep{derosa14}. The \feii\ template used here is the combination of templates from \cite{vestergaard01} and \cite{tsuzuki06}. \citet[][hereafter VW01]{vestergaard01} constructed an empirical ultraviolet iron template covering the wavelength range 1250--3090 \AA\ based on spectra of the narrow line Seyfert 1 galaxy I Zw 1. At that time, the iron emission underlying the \mgii\ line could not be well estimated, so the iron emission in the template was set to zero over this region. \citet[][hereafter T06]{tsuzuki06} derived an \feii\ template from the spectrum of I Zw 1 for the regions 2200--3500 and 4200--5600 \AA\ and used synthetic spectra calculated with the CLOUDY photoionization code to separate the underlying \feii\ emission from the \mgii\ emission line. 
In this work, we combine these two templates by combining the \feii\ emission from the VW01 template for 1100--2200 \AA\ and the T06 template for 2200--3500 \AA, in order to obtain a template covering a wide wavelength range and also containing the \feii\ emission beneath the \mgii\ line. When fitting a spectrum, the iron template is broadened by convolving the template with a Gaussian kernel derived from the width the of \mgii\ line.

Gaussian fits of the \civ\ and \mgii\ emission lines are then performed on the continuum-subtracted spectrum. 
The \ion{Si}{4} and \ion{C}{3}] lines are also fitted if they are visible. However, the \ion{Si}{4} lines sometimes are too close to the edge of the recorded spectrum and thus have lower S/N. Most of the \ion{C}{3}] lines are fully or partly located within the region affected by strong telluric absorption (i.e., $\sim$ 13500--14200 \AA). So the fitting of these two lines has lower quality than the fitting of the \civ\ and \mgii\ lines, and we are not using these two lines for scientific analysis in this paper. For most cases, a two-component Gaussian profile is used to fit each emission line, while for a few objects only a one-component Gaussian is used. 
For example, for the quasar J0910--0414, we use only one Gaussian to fit its \mgii\ line due to multiple strong absorption features around the \mgii\ line. When masking all the absorption features, the wide range of absorptions results in a significant gap at the line center, such that a two-Gaussian model fit would result in a double-peaked emission line model. For a similar reason, a one-component Gaussian model is also used for the \civ\ line of the quasar J1058+2930. Four quasars do not have \civ\ fitting, including the quasar J0525--2406 that has no spectrum blueward of 14500 \AA\ and the other three quasars with unusual spectral features (i.e., J0910--0414, J1316+1028, and J1535+1943; see details in the next subsection).
The redshifts derived from the \civ\ and \mgii\ emission lines are based on the line centroids \citep{peterson04} rather than the line peaks. In this case,  any strong blue/redshifted component will result in a different redshift measurement from the measurement using the line peak.
The uncertainties of all spectral measurements are estimated using a Monte Carlo approach, following \cite{yang20a} \citep[also ][]{shen19, wang20}.
For each spectrum, we generate 100 mock spectra by randomly adding Gaussian noise at each pixel with standard deviation equal to the spectral error at that pixel. We apply the same fitting procedure to each mock spectrum to obtain the corresponding measurements. Then the uncertainties of the spectral measurements are estimated as the average of the 16\% and 84\% percentile deviations from the median value. 
The best fits of the continuum and the \civ\ and \mgii\ lines for each quasar are shown in Figures \ref{fig:fitting} and \ref{fig:allspec01}.

The spectral fitting yields a set of spectral properties of these quasars, including continuum slope, luminosity, emission line FWHM, and line rest-frame equivalent width (EW). The redshifts derived from the UV emission lines and line velocity shifts will be  discussed in detail in Section 5.1. 
The quasars in our sample have power-law continuum slopes in the range of --1.74 to --0.24 ($f_{\rm \lambda}\propto \lambda^{\alpha_{\rm \lambda}}$), with a mean of $\alpha_{\rm \lambda} = -1.2$ and a 1$\sigma$ dispersion of 0.4. 
The mean is in good agreement with the mean slopes from quasar samples at similar redshifts (e.g., $\alpha_{\rm \lambda}=-1.2$ from the NIR spectral sample in \citealt{mazzucchelli17} and $\alpha_{\rm \lambda}=-1.4$ in \citealt{schindler20}). It is also consistent with the quasar composites generated from low-redshift quasars (e.g., $\alpha_{\rm \lambda}=-1.5$ in \citealt{vandenberk01} and $\alpha_{\rm \lambda}=-1.7$ in \citealt{selsing16}) within the uncertainty, although our result has a slightly redder slope.
The absolute rest-frame 1450 \AA\ magnitudes are derived from the best-fit power law continuum directly. We also measure the rest-frame 3000 \AA\ luminosity and convert it to a bolometric luminosity assuming a bolometric correction factor of 5.15 \citep{richards06,shen11}. These quasars are in the luminosity range 0.5 -- 3.4 $\times 10^{47}$ erg s$^{-1}$. The range of FWHM of the \civ\ lines is $\sim$ 1900 -- 12000 km\,s$^{-1}$ with a mean of 5900 km\,s$^{-1}$, and the \mgii\ lines have FWHMs of $\sim$ 1700 -- 5500 km\,s$^{-1}$ with a mean of 3000 km\,s$^{-1}$.  The EWs of \civ\ are in the range of 6 to 70 \AA\ and have a mean of 30 \AA. The EWs of the \mgii\ line are from 8 to 35 \AA\ with a mean of 20 \AA. All these measurements are summarized in Table \ref{tab:fitting}. 

\subsection{Notes on Individual Objects}
Some quasars have unusual spectral features such as reddened continuum shapes or strong absorption lines. We describe these objects and their spectral fitting separately. 

{\it BAL quasars -- }
 In our sample, there are nine quasars with significant BAL features (see details in Section 6.2), which significantly affect the spectral fitting. In particular, the quasar J0910--0414 has multiple metal absorptions within the emission-line profiles in addition to its BAL feature, indicative of both outflow and inflow, which will be discussed in a separate paper. Its strong absorption features mask most of the emission at $< 1570$ \AA\ (rest-frame), therefore we use only the longer wavelength range for spectral fitting. For the quasars J0246--5219 and J0038--1527, we also mask the wavelength range shorter than rest-frame 1500 \AA\ because of their strong BAL absorptions on the blue side. For the other BAL quasars, we mask the BAL troughs for spectral fitting. 
 
{\it Red quasars --}
The quasars J0246--5219, J0319--1008, and J1316+1028 have red continua with slopes $\alpha_{\rm \lambda} > -0.5$. They all have red $J$--W1 colors ($J$--W1 $> 2.5$). J1316+1028's spectrum does not show a \civ\ line and has only a tentative \ion{C}{3}] line, which may be affected by the low S/N in this wavelength range. This quasar shows a strong BAL feature in its observed optical spectrum \citep{wang19} but much weaker absorption features in the NIR spectrum.

{\it Unusual reddened quasar -- }
The quasar J1535+1943 has a reddened continuum in $Y$ and $J$ but a relatively blue continuum at redder wavelengths. If this reddening is caused by dust extinction, the extinction, relatively flat at wavelengths redward of rest-frame $1700$ \AA\ and steeply rising at shorter wavelengths, is quite similar to the dust extinction of quasar SDSS1048+46 at $z=6$ described in \cite{maiolino04}. This kind of dust extinction detected in high-redshift quasar spectra could be evidence for the origin of early dust formation (e.g., a supernova origin for the dust). A detailed discussion of its dust extinction will be presented in a subsequent paper (J. Yang et al. in prep).
Given the relatively flat extinction at $> 1700$ \AA, we fit its continuum using only the spectrum at longer wavelengths. A slope of $\alpha_{\rm \lambda}=-0.92\pm0.02$ is obtained. Due to the uncertain dust extinction, the luminosity measured from the observed spectrum is a lower limit, and thus its BH mass is also a lower limit.

\begin{deluxetable*}{l l l l l l l l l l l l l}
\rotate 
\tablecaption{Spectral Fitting and Quasar Properties}
\tabletypesize{\footnotesize}
\tablewidth{0pt}
\tablehead{
\colhead{Name} &
\colhead{$z_{\rm [CII]}$} &
\colhead{$z_{\rm CIV}$} &
\colhead{$z_{\rm MgII}$} &
\colhead{$M_{\rm 1450}$} &
\colhead{FWHM$_{\rm CIV}$ } &
\colhead{FWHM$_{\rm MgII}$ } &
\colhead{EW$_{\rm CIV}$ } &
\colhead{EW$_{\rm MgII}$ } &
\colhead{$L_{\rm Bol}$} &
\colhead{$M_{\rm BH}$} &
\colhead{$\alpha_{\rm \lambda}$\tablenotemark{a}} &
\colhead{$\lambda_{\rm Edd}$}
\\
\nocolhead{} & \nocolhead{} & \nocolhead{} & \nocolhead{} & \nocolhead{} & \colhead{(km s$^{-1}$)} & \colhead{(km s$^{-1}$)} & \colhead{(\AA)} & \colhead{(\AA)} & \colhead{($10^{46} $\rm erg s$^{-1}$)} & \colhead{($10^{9}\,M_{\odot}$)}  &  \nocolhead{} & \nocolhead{}
}
\startdata
J0024+3913  &  6.621$\pm$0.002  &  6.608$\pm$0.002  &  6.620$\pm$0.004  &  --25.65  &  1908$\pm$26  &  1741$\pm$118  &  68.8$\pm$11.1  &  28.8$\pm$6.7  &  7.8$\pm$1.0  &  0.27$\pm$0.02  &  --1.11$\pm$0.02  &  2.3$\pm$0.4 \\
J0038--1527\tablenotemark{c}  &  7.0340$\pm$0.0003  &  6.929$\pm$0.003  &  6.999$\pm$0.001  &  --27.13  &  7800$\pm$349  &  2954$\pm$17  &  12.0$\pm$1.5  &  15.0$\pm$0.6  &  23.7$\pm$0.9  &  1.36$\pm$0.05  &  --1.46$\pm$0.02  &  1.4$\pm$0.1 \\
J0045+0901  &  6.4694$\pm$0.0025  &  6.43$\pm$0.02  &  6.441$\pm$0.004  &  --25.86  &  6373$\pm$2299  &  2816$\pm$110  &  12.6$\pm$2.8  &  15.5$\pm$3.6  &  6.2$\pm$0.6  &  0.63$\pm$0.02  &  --1.69$\pm$0.03  &  0.8$\pm$0.1 \\
J0218+0007  &  6.7700$\pm$0.0013  &  6.725$\pm$0.007  &  6.766$\pm$0.004  &  --25.55  &  5406$\pm$983  &  2745$\pm$73  &  18.0$\pm$11.4  &  26.0$\pm$6.4  &  6.4$\pm$1.4  &  0.61$\pm$0.07  &  --1.24$\pm$0.13  &  0.8$\pm$0.2 \\
J0246--5219\tablenotemark{c}  &  6.8876$\pm$0.0003  &  6.851$\pm$0.002  &  6.86$\pm$0.02  &  --25.36  &  4070$\pm$286  &  3211$\pm$523  &  30.6$\pm$1.8  &  30.4$\pm$18.2  &  10.2$\pm$1.0  &  1.05$\pm$0.37  &  --0.37$\pm$0.05  &  0.8$\pm$0.3 \\
J0252--0503  &  7.0006$\pm$0.0009  &  6.867$\pm$0.005  &  6.99$\pm$0.02  &  --26.63  &  11286$\pm$698  &  3327$\pm$126  &  17.3$\pm$1.0  &  17.6$\pm$2.8  &  13.2$\pm$0.4  &  1.28$\pm$0.09  &  --1.62$\pm$0.02  &  0.8$\pm$0.1 \\
J0313--1806\tablenotemark{c}  &  7.6423$\pm$0.0013  &  7.523$\pm$0.01  &  7.611$\pm$0.004  &  --26.13  &  8740$\pm$1828  &  3670$\pm$405  &  14.2$\pm$0.9  &  9.5$\pm$2.7  &  14.0$\pm$0.7  &  1.61$\pm$0.40  &  --0.91$\pm$0.01  &  0.7$\pm$0.2 \\
J0319--1008  &  6.8275$\pm$0.0021  &  6.809$\pm$0.005  &  6.816$\pm$0.004  &  --25.36  &  3164$\pm$205  &  2006$\pm$20  &  70.1$\pm$26.2  &  32.8$\pm$4.8  &  9.6$\pm$1.4  &  0.40$\pm$0.03  &  --0.45$\pm$0.35  &  1.9$\pm$0.3 \\
J0411--0907  &  6.8260$\pm$0.0007  &  6.790$\pm$0.005  &  6.827$\pm$0.006  &  --26.58  &  4046$\pm$1047  &  2729$\pm$96  &  43.2$\pm$4.8  &  19.6$\pm$1.6  &  15.9$\pm$1.0  &  0.95$\pm$0.09  &  --1.31$\pm$0.03  &  1.3$\pm$0.2 \\
J0439+1634\tablenotemark{b}\tablenotemark{c}  &  6.5188$\pm$0.0004  &  6.492$\pm$0.002  &  6.519$\pm$0.003  &  --25.31  &  6067$\pm$277  &  3030$\pm$65  &  41.0$\pm$1.8  &  17.8$\pm$0.6  &  4.6$\pm$0.1  &  0.63$\pm$0.02  &  --1.41$\pm$0.03  &  0.6$\pm$0.1 \\
J0525--2406  &  6.5397$\pm$0.0001  &  ---  &  6.543$\pm$0.002  &  --25.47  &  ---  &  1877$\pm$345  &  ---  &  14.5$\pm$11.9  &  6.8$\pm$3.5  &  0.29$\pm$0.04  &  --1.07$\pm$0.93  &  1.8$\pm$1.0 \\
J0706+2921\tablenotemark{c}  &  6.6037$\pm$0.0003  &  6.54$\pm$0.02  &  6.5925$\pm$0.0004  &  --27.44  &  8673$\pm$467  &  3372$\pm$106  &  30.5$\pm$4.2  &  20.4$\pm$1.5  &  33.9$\pm$1.5  &  2.11$\pm$0.16  &  --1.35$\pm$0.03  &  1.3$\pm$0.1 \\
J0803+3138  &  ---  &  6.332$\pm$0.005  &  6.384$\pm$0.004  &  --26.49  &  8073$\pm$743  &  3460$\pm$173  &  23.2$\pm$2.3  &  18.5$\pm$3.4  &  13.4$\pm$1.1  &  1.40$\pm$0.18  &  --1.43$\pm$0.04  &  0.8$\pm$0.1 \\
J0829+4117  &  ---  &  6.736$\pm$0.001  &  6.773$\pm$0.007  &  --26.07  &  4405$\pm$1411  &  2488$\pm$107  &  63.7$\pm$9.2  &  21.7$\pm$3.6  &  12.8$\pm$1.2  &  0.71$\pm$0.02  &  --0.95$\pm$0.02  &  1.4$\pm$0.1 \\
J0837+4929  &  ---  &  6.677$\pm$0.002  &  6.702$\pm$0.001  &  --26.33  &  4165$\pm$401  &  2577$\pm$36  &  37.8$\pm$2.6  &  34.5$\pm$1.9  &  14.5$\pm$0.4  &  0.81$\pm$0.01  &  --1.11$\pm$0.03  &  1.4$\pm$0.1 \\
J0839+3900\tablenotemark{c}  &  ---  &  6.86$\pm$0.01  &  6.9046$\pm$0.0003  &  --26.36  &  5904$\pm$258  &  2233$\pm$21  &  15.3$\pm$1.4  &  16.6$\pm$1.1  &  17.8$\pm$0.7  &  0.671$\pm$0.003  &  --0.87$\pm$0.01  &  2.1$\pm$0.1 \\
J0910--0414\tablenotemark{c}  &  6.6363$\pm$0.0003  &  ---  &  6.610$\pm$0.003  &  --26.61  &  ---  &  5396$\pm$544  &  ---  &  15.3$\pm$1.2  &  15.0$\pm$1.1  &  3.59$\pm$0.61  &  --1.43$\pm$0.01  &  0.3$\pm$0.1 \\
J0910+1656  &  6.7289$\pm$0.0005  &  6.718$\pm$0.003  &  6.719$\pm$0.005  &  --25.34  &  2181$\pm$341  &  2358$\pm$28  &  61.1$\pm$11.6  &  30.2$\pm$5.3  &  5.3$\pm$0.6  &  0.41$\pm$0.03  &  --1.22$\pm$0.09  &  1.0$\pm$0.1 \\
J0921+0007  &  6.5646$\pm$0.0003  &  6.553$\pm$0.005  &  6.5654$\pm$0.0002  &  --25.19  &  2221$\pm$154  &  1813$\pm$14  &  43.5$\pm$7.6  &  20.8$\pm$2.1  &  6.1$\pm$0.6  &  0.26$\pm$0.01  &  --0.86$\pm$0.09  &  1.9$\pm$0.2 \\
J0923+0402\tablenotemark{c}  &  6.6330$\pm$0.0003  &  6.59$\pm$0.02  &  6.612$\pm$0.002  &  --26.68  &  4680$\pm$531  &  3454$\pm$109  &  16.0$\pm$2.5  &  16.4$\pm$3.2  &  21.7$\pm$3.0  &  1.77$\pm$0.02  &  --1.00$\pm$0.07  &  1.0$\pm$0.1 \\
J0923+0753  &  6.6817$\pm$0.0005  &  6.652$\pm$0.01  &  6.682$\pm$0.002  &  --25.5  &  4099$\pm$526  &  2640$\pm$682  &  40.8$\pm$22.6  &  34.7$\pm$15.6  &  4.9$\pm$2.0  &  0.49$\pm$0.15  &  --1.55$\pm$0.06  &  0.8$\pm$0.4 \\
J1007+2115  &  7.5149$\pm$0.0004  &  7.39$\pm$0.04  &  7.48$\pm$0.01  &  --26.73  &  7988$\pm$2045  &  3152$\pm$168  &  10.0$\pm$1.6  &  20.1$\pm$6.1  &  20.4$\pm$1.3  &  1.43$\pm$0.22  &  --1.14$\pm$0.01  &  1.1$\pm$0.2 \\
J1058+2930  &  6.5846$\pm$0.0005  &  6.523$\pm$0.002  &  6.585$\pm$0.005  &  --25.68  &  5709$\pm$128  &  2656$\pm$97  &  32.5$\pm$11.7  &  19.0$\pm$6.8  &  5.8$\pm$1.5  &  0.54$\pm$0.03  &  --1.57$\pm$0.07  &  0.8$\pm$0.2 \\
J1104+2134  &  6.7662$\pm$0.0009  &  6.739$\pm$0.002  &  6.766$\pm$0.005  &  --26.63  &  6396$\pm$1242  &  3695$\pm$225  &  32.8$\pm$4.6  &  24.8$\pm$3.0  &  15.1$\pm$0.9  &  1.69$\pm$0.15  &  --1.44$\pm$0.04  &  0.7$\pm$0.1 \\
J1120+0641  &  7.0851$\pm$0.0005  &  7.016$\pm$0.002  &  7.070$\pm$0.003  &  --26.44  &  8101$\pm$281  &  3402$\pm$73  &  25.9$\pm$2.4  &  20.9$\pm$2.2  &  13.4$\pm$1.0  &  1.35$\pm$0.04  &  --1.36$\pm$0.02  &  0.8$\pm$0.1 \\
J1129+1846  &  ---  &  6.804$\pm$0.008  &  6.824$\pm$0.001  &  --25.73  &  3008$\pm$997  &  1774$\pm$36  &  30.1$\pm$17.6  &  17.6$\pm$3.6  &  8.4$\pm$1.9  &  0.29$\pm$0.02  &  --1.10$\pm$0.26  &  2.3$\pm$0.5 \\
J1135+5011  &  6.5851$\pm$0.0008  &  6.53$\pm$0.01  &  6.579$\pm$0.001  &  --26.16  &  7469$\pm$397  &  3762$\pm$129  &  32.4$\pm$5.9  &  22.0$\pm$2.7  &  10.8$\pm$0.8  &  1.49$\pm$0.05  &  --1.30$\pm$0.02  &  0.6$\pm$0.1 \\
J1216+4519  &  ---  &  6.56$\pm$0.02  &  6.648$\pm$0.003  &  --25.57  &  8947$\pm$410  &  2816$\pm$292  &  34.3$\pm$10.2  &  20.0$\pm$3.4  &  5.8$\pm$1.2  &  0.61$\pm$0.20  &  --1.40$\pm$0.01  &  0.8$\pm$0.3 \\
J1316+1028\tablenotemark{c}  &  ---  &  ---  &  6.329$\pm$0.005  &  --25.67  &  ---  &  2866$\pm$763  &  ---  &  7.9$\pm$3.0  &  14.8$\pm$3.3  &  1.01$\pm$0.37  &  --0.24$\pm$0.39  &  1.2$\pm$0.5 \\
J1342+0928  &  7.5413$\pm$0.0007  &  7.37$\pm$0.02  &  7.51$\pm$0.01  &  --26.67  &  11989$\pm$1236  &  2640$\pm$215  &  12.9$\pm$1.4  &  13.9$\pm$5.8  &  13.3$\pm$1.1  &  0.81$\pm$0.18  &  --1.67$\pm$0.04  &  1.3$\pm$0.3 \\
J1535+1943  &  ---  &  ---  &  6.370$\pm$0.001  &  --27.09  &  ---  &  4372$\pm$266  &  ---  &  13.4$\pm$1.3  &  33.5$\pm$1.7  &  3.53$\pm$0.33  &  --0.92$\pm$0.02  &  0.8$\pm$0.1 \\
J1724+1901  &  ---  &  6.45$\pm$0.03  &  6.480$\pm$0.001  &  --25.55  &  3716$\pm$1267  &  2704$\pm$62  &  7.1$\pm$5.5  &  11.3$\pm$3.2  &  8.4$\pm$1.3  &  0.67$\pm$0.08  &  --0.88$\pm$0.14  &  1.0$\pm$0.2 \\
J2002--3013  &  6.6876$\pm$0.0004  &  6.64$\pm$0.02  &  6.673$\pm$0.001  &  --26.9  &  7298$\pm$1005  &  3598$\pm$351  &  8.4$\pm$3.2  &  17.9$\pm$3.0  &  15.4$\pm$1.9  &  1.62$\pm$0.27  &  --1.74$\pm$0.04  &  0.8$\pm$0.2 \\
J2102--1458  &  6.6645$\pm$0.0002  &  6.611$\pm$0.009  &  6.652$\pm$0.003  &  --25.53  &  6146$\pm$295  &  3083$\pm$186  &  23.2$\pm$4.4  &  20.1$\pm$3.0  &  6.0$\pm$0.5  &  0.74$\pm$0.11  &  --1.31$\pm$0.04  &  0.6$\pm$0.1 \\
J2211--6320  &  6.8449$\pm$0.0003  &  6.73$\pm$0.01  &  6.83$\pm$0.01  &  --25.38  &  7985$\pm$394  &  2679$\pm$608  &  26.2$\pm$4.4  &  16.1$\pm$16.5  &  5.9$\pm$0.2  &  0.55$\pm$0.24  &  --1.17$\pm$0.10  &  0.8$\pm$0.4 \\
J2232+2930  &  6.666$\pm$0.004  &  6.671$\pm$0.007  &  6.655$\pm$0.003  &  --26.26  &  3876$\pm$178  &  5504$\pm$159  &  38.4$\pm$8.8  &  27.0$\pm$7.2  &  10.0$\pm$1.7  &  3.06$\pm$0.36  &  --1.53$\pm$0.04  &  0.3$\pm$0.1 \\
J2338+2143  &  ---  &  6.49$\pm$0.03  &  6.565$\pm$0.009  &  --26.0  &  2296$\pm$3206  &  2516$\pm$113  &  5.9$\pm$4.1  &  14.9$\pm$5.9  &  7.6$\pm$1.3  &  0.56$\pm$0.03  &  --1.57$\pm$0.03  &  1.1$\pm$0.2 \\
 \enddata
\tablenotetext{a}{Continuum slope $\alpha_{\rm \lambda}$ ($f_{\lambda} \propto \lambda^{\alpha_{\rm \lambda}}$).}
\tablenotetext{b}{The measurements of the quasar J0439+1634 have been corrected for gravitational lensing using a magnification of 51.3 \citep{fan19}.}
\tablenotetext{c}{BAL quasars. See details in Section 6.2.}
 \label{tab:fitting}
\end{deluxetable*}

\section{Black Hole Mass and Eddington Ratio}
In this section, we report the measurements of BH masses based on the NIR spectra. Together with bolometric luminosities measured from spectral fitting, we then estimate the Eddington ratios of these quasars and compare them with quasar samples at both lower and similar redshift ranges.

\subsection{Virial Black Hole Mass}
Assuming virial motion for line-emitting gas in the quasar BLR and based on the correlation between the measured BLR size and quasar continuum luminosity (i.e., the $R-L$ relation), quasar BH masses can be estimated from single-epoch spectra by measuring the line width of the UV and optical broad emission lines and continuum luminosity. Emission lines including \civ, \mgii, H$\alpha$, and H$\beta$ have all been used for virial BH mass estimators \cite[e.g.,][]{mclure02, mclure04, vestergaard02, vestergaard06, greene05, shen11}. Calibration coefficients used in the BH mass estimators are determined using  samples with mass measurements based on reverberation mapping (RM) at low redshifts. 
The uncertainty of this method is estimated to be on the order of $\sim$ 0.5 dex, inferred from the residuals in the calibrations against RM-based BH masses \cite[e.g.,][]{mclure02,vestergaard06,shen13}.
Both RM-based measurements and comparison of single-epoch mass estimators have demonstrated that H$\beta$ is the most reliable among the emission lines typically used for virial BH mass estimation \citep[e.g.,][]{shen13}. However, at $z>4$, the H$\beta$ line moves outside the NIR spectral coverage. Given that the \civ\ line has been suggested to be associated with outflows, particularly considering the large blueshift of the \civ\ line found in high-redshift quasars \citep[e.g.,][]{meyer19}, the \mgii\ line is more acceptable for use as a BH mass estimator at high redshift. The current \mgii-based scaling relation is calibrated based on the H$\beta$ relations derived from reverberation mapping \citep[e.g.,][]{vestergaard09}. 

\begin{figure}
\centering 
\epsscale{1.2}
\plotone{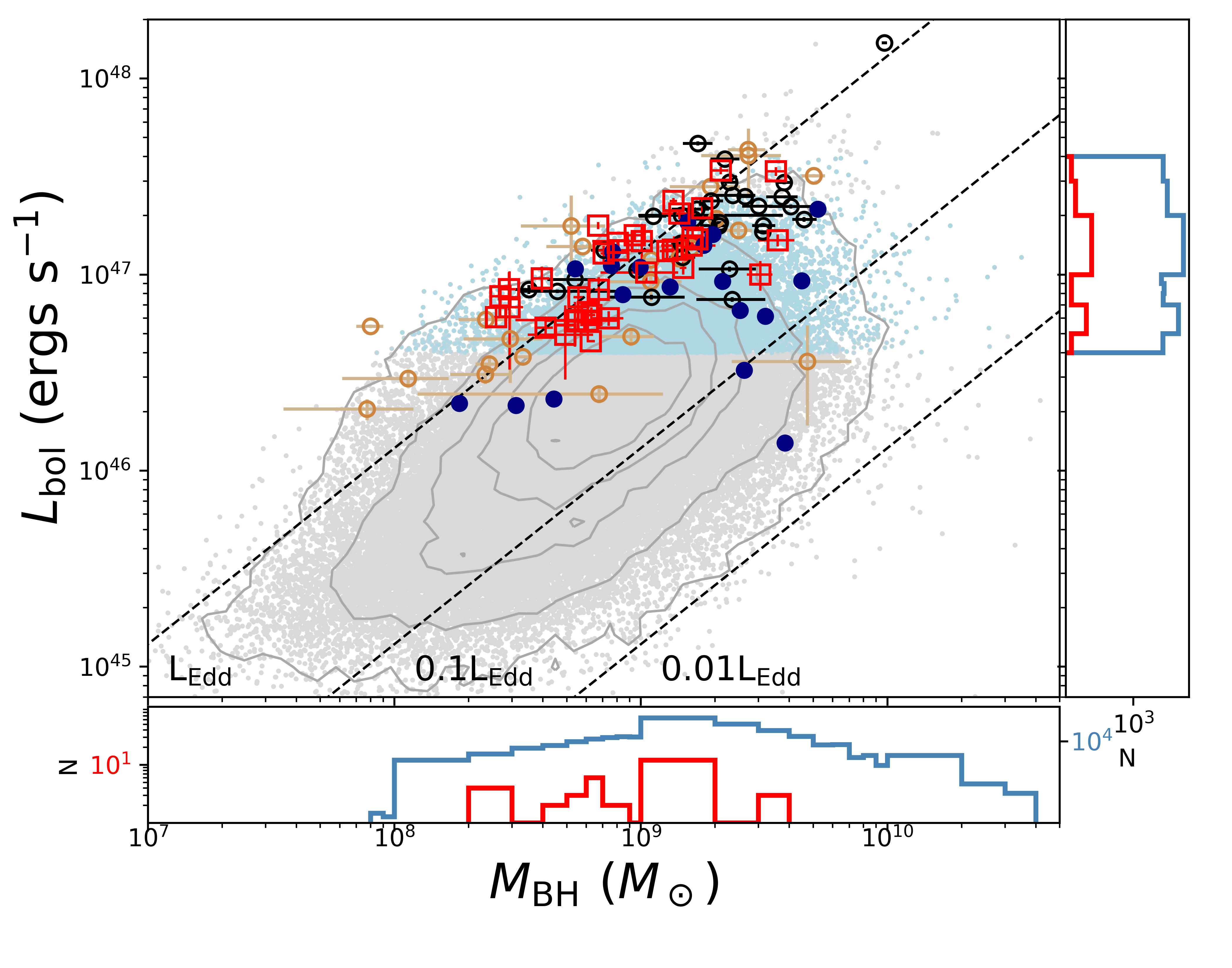} 
\plotone{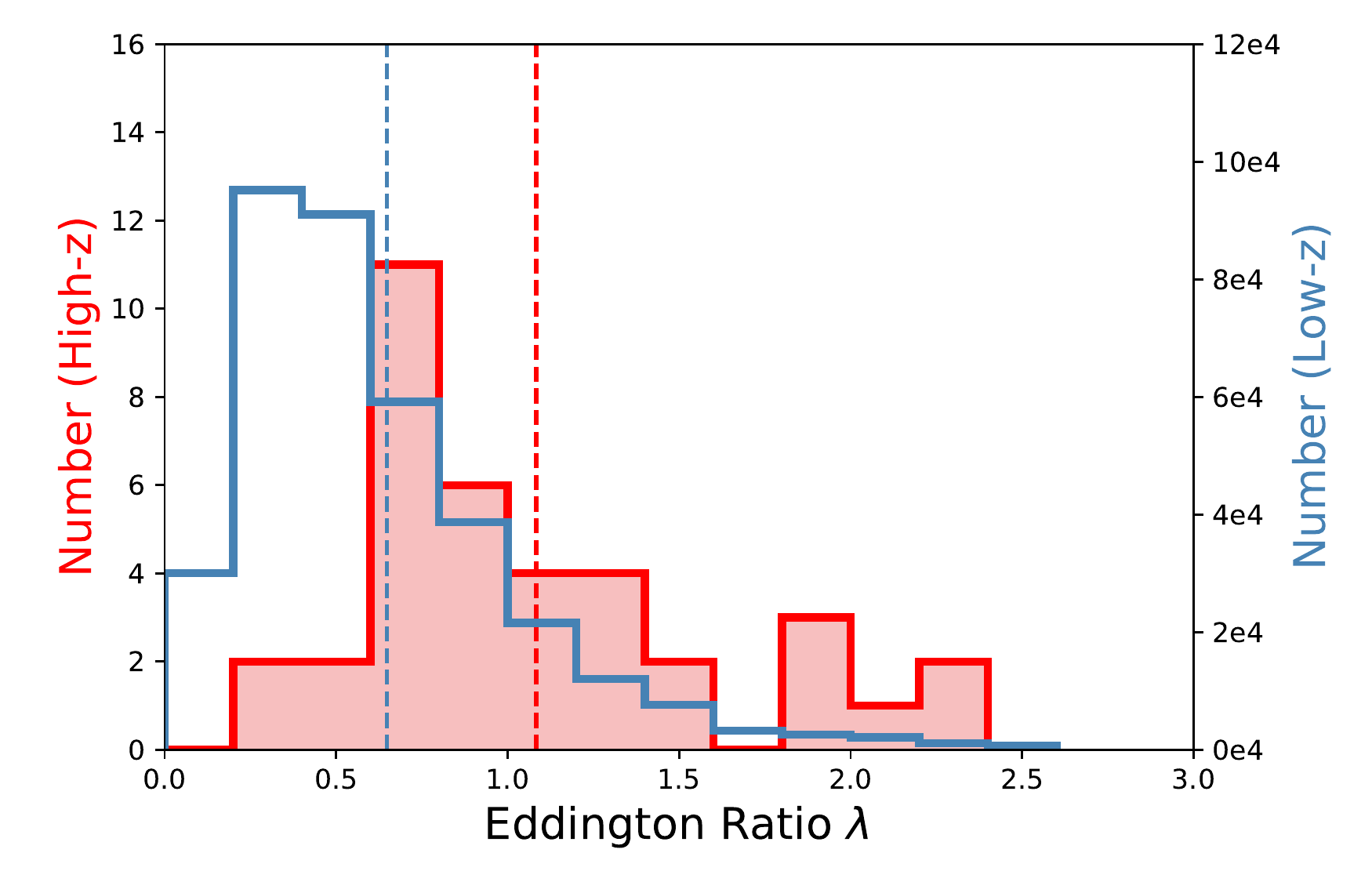} 
\caption{{\bf Top:} New measurements of the quasar bolometric luminosities and BH masses of our sample (red open squares), compared with measurements of $z\sim6$ quasars \citep{willott03,kurk07,willott10,derosa11,mazzucchelli17,shen19,schindler20} and the SDSS lower redshift sample \citep[][gray dots]{shen11}, estimated based on the same BH mass estimator. The black points are measurements from \cite{schindler20}, and the results from \cite{shen19} are shown in dark blue. Measurements from other works \citep{willott03,kurk07,willott10,derosa11,mazzucchelli17} are shown in light brown. Quasars duplicated between these samples are excluded. The light blue dots represent a luminosity-matched sample selected from the SDSS lower redshift sample. The histogram towards the right shows the luminosity distributions of the luminosity-matched low-redshift sample (blue) and our sample (red). The bottom histogram shows the BH mass distributions of the low-redshift sample (blue) and our sample (red). 
{\bf Bottom:} The distribution of Eddington ratios ($\lambda = L_{\rm bol}/L_{\rm Edd}$) measured from quasars in our sample (red), compared with the luminosity-matched low-redshift quasar sample described above (blue). The Eddington ratios of quasars in our sample peak at $\lambda \sim 0.8$ and have a mean of 1.09 (red dashed line), while the low-redshift sample has a mean Eddington ratio of 0.65.} \label{fig:bh_l_edd}
\end{figure}

We estimate the BH masses of our quasars based on the continuum luminosity at 3000 \AA\ (rest-frame) and the FWHM of the \mgii\ line, by adopting the empirical relation from \citet[][VO09]{vestergaard09}:
\begin{equation}
\small
\frac{M_{\rm BH}}{M_\odot} = 10^{6.86} \left[\frac{\lambda L_{\lambda}\rm{(3000~\AA)}}{\rm{10^{44}~ erg~s^{-1}}}\right]^{0.5}  \left[\frac{\rm{FWHM_{(Mg \ II)}}}{\rm{1000~km~s^{-1}}}\right]^2
\end{equation}
The systematic uncertainties of this scaling relation could be up to $\sim0.55$ dex (VO09). The BH mass uncertainties reported in this paper are estimated from spectral fitting only and do not include the systematic uncertainties.
We derive BH masses of all 37 quasars in this sample and find them to be in the range $2.6 \times10^{8} -3.6\times10^{9}\,M_{\odot}$. The individual measurements are listed in Table \ref{tab:fitting}.
For our sample, if we use the BH mass estimator from \citet[][MD04]{mclure04}, we obtain $\sim$ 0.76 to 0.96 times smaller BH masses, dependent on the luminosity of the quasar. 
If we use the estimator from \citet[][S11]{shen11}, the BH masses of these quasars change to $\sim$ 1.30 to 1.65 times larger.

The \civ\ line has also been used to estimate the BH mass \citep[e.g.,][]{vestergaard06,coatman17}. Since it is thought to include components with non-virial origins, corrections have been suggested for \civ\ single-epoch BH mass estimators \citep[e.g.,][]{park13,coatman16,coatman17,zuo20}. We apply the scaling relation from \cite{vestergaard06} and the empirical correction as a function of \civ\ blueshift derived from low-redshift quasars in \cite{coatman17} to estimate \civ-based BH masses. For the quasars in our sample, the ratios of \civ\ BH masses to \mgii\ BH masses have a large scatter, from 0.3 to 3.9 (with a mean of 2.0). Note that the redshifts used here for estimating \civ\ blueshifts are [\cii] or \mgii\ redshifts, while the empirical correction is based on H$\alpha$ line redshifts. In addition, the quasars in our sample show larger \civ\ blueshifts compared with low-redshift quasars (see Section 5.1), which will increase the uncertainty of \civ\ BH masses for these high-redshift quasars.
Therefore, we will not directly compare the \civ\ and \mgii\ BH masses for individual objects and adopt only the \mgii-based BH masses in subsequent analysis.

The BH masses of these quasars are plotted in Figure \ref{fig:bh_l_edd} (top) together with their bolometric luminosities, compared with the measurements for other $z\gtrsim6$ quasars \citep{willott03,kurk07,willott10,derosa11,mazzucchelli17,shen19, schindler20} and the low-redshift SDSS quasar sample \citep{shen11}. All BH masses used here are derived from the same BH mass estimator. Figure \ref{fig:bh_l_edd} shows that the quasars in our sample are located close to the line of Eddington luminosity, indicating that these SMBHs are accreting close to the Eddington limit, similar to the behavior found from most of other known $z \gtrsim 6$ quasars.  
A sample of low-redshift quasars is selected as comparison from the SDSS DR7 quasar catalog with \mgii-based BH masses \citep{shen11}. We select quasars in the redshift range $0.4 \le z \le 2.1$ to ensure sufficient spectral coverage of the entire \mgii\ line in the SDSS DR7 spectra. We adopt the BH masses from \citep{shen11}, derived using the same estimator (i.e., VO09). A luminosity-matched control sample is also selected from them to compare the BH masses of low- and high-redshift samples with the same luminosity distribution. We select quasars from the DR7 sample following the luminosity distribution of our sample, by randomly choosing 10 times more quasars than the quasars in our sample at each bolometric luminosity bin and repeating the sampling 1000 times.
Figure \ref{fig:bh_l_edd} shows that the low-redshift control sample and our sample are consistent with being drawn from the same the same luminosity distribution, according to a two-sample K-S test \citep[][$p=0.9$]{kolmogorov1933}, while their BH masses are from different distributions with $p \ll$ 0.01 and low-redshift quasars have more massive BHs.
We can also quantify this difference by comparing their Eddington ratios.

\subsection{Eddington Ratio}
Based on the bolometric luminosities and BH masses measured above, we derive the Eddington ratios. Note that the uncertainty of the Eddington ratio derived based on the \mgii-based BH mass is subject to the same systematic uncertainty as the BH mass. As shown in Figure \ref{fig:bh_l_edd} (bottom), the Eddington ratios of these high-redshift quasars span values from 0.26 to 2.3, with a mean of 1.08 (a median value of 0.85) and a peak at $\lambda_{\rm Edd} \sim 0.8$. There are 16 quasars with Eddington ratios higher than one, including three quasars with $\lambda_{\rm Edd} >2$. The radio-loud quasar J1129+1846 and the quasar J0024+3913 have already been reported as super-Eddington quasars in \cite{banados21} and \cite{wang21a}, respectively. The quasar J0839+3900 is the third one and has $\lambda_{\rm Edd} = 2.1$. The Eddington ratios for individual quasars are listed in Table \ref{tab:fitting}.

The high average Eddington ratio has been reported in a number of previous works on quasar \mgii-based BH masses at redshifts $z > 5.8$ \citep[e.g.,][]{kurk07, jiang07, willott10, schindler20, yang20a, wang21b}, with Eddington ratios close to unity. The question of whether the high Eddington ratios of these high-redshift quasars are intrinsic or affected by selection effects is still debated. Compared to lower redshift quasar samples, the known high-redshift quasars are located at the relatively luminous end of the distribution, as a result of limited survey depth. The correlations among luminosity, BH mass, and Eddington ratio complicate the determinations of the Eddington ratio distribution based on flux-limited quasar samples, and may limit our study of high-redshift quasars to a relatively narrow range of BH mass and Eddington ratio.

It has been suggested that the high Eddington ratios in some high-redshift quasar samples could be due to the high luminosities of quasars in those samples. 
If these most luminous quasars ($L_{\rm bol} \sim 10^{47}$ erg s$^{-1}$) accrete at $\lambda_{\rm Edd} \sim$ 0.1, they would require a BH mass of $\sim 10^{10}\,M_{\odot}$. 
In this case, low-luminosity quasars would help to overcome the selection bias. \cite{willott10} report the observations of nine faint CFHQS quasars at $z\sim6$ and find high Eddington ratios in these quasars, with a median of $\lambda=1.2$. This result therefore indicates that BHs with highly active status exist in both high and low-luminosity quasar populations at $z \gtrsim 6$. The six faint quasars from the HSC quasar survey have a wider range of Eddington ratios from 0.16 to 1.1 \citep{matsuoka19b, onoue19}, while the least luminous quasar in this sample, which is also the least massive SMBH at $z>5.8$, has a Eddington ratio of 1.1. The results from the CFHQS and HSC samples suggest a broad range of the Eddington ratio distribution of less luminous $z\gtrsim 6$ quasars.
The recent measurements of 50 $z\sim6$ luminous quasars in \cite{shen19}, however, give low Eddington ratios with a median of 0.3. 
\cite{shen19} check objects overlapping between their sample and earlier works and suggest that the apparent difference in their reported Eddington ratios for the common objects is largely due to the difference in the adopted BH mass estimators.

We compare our measurements with the Eddington ratios of low-redshift quasars using the luminosity-matched control sample described above. The low-redshift quasars in the control sample have a mean Eddington ratio of 0.65 (a median of 0.53) and a peak at $\sim 0.3$, which are significantly lower than the values in our high-redshift sample. We also note that the luminosity-matched low-redshift quasar samples in this work and in \cite{shen19} are all from the SDSS DR7 quasar properties catalog \citep{shen11} and the high-redshift quasars in both works have similar luminosity ranges, while the low-redshift luminosity-matched sample used in this work has significantly higher Eddington ratios (median $\sim 0.5$) than that in \cite{shen19} (median $\sim 0.3$). The main reason for this discrepancy is the different BH mass estimators, as we are using VO09 but \cite{shen19} use S11. The measurements in \cite{shen11} derived from the other two estimators (MD04 and S11) result in lower Eddington ratios for the control sample, with median values of $\sim 0.3-0.4$.
As described above, if using the MD04 BH mass estimator, for our high-redshift quasars, we will have smaller BH masses and thus higher Eddington ratios, with a median value of 1.08. The S11 estimator will yield a median Eddington ratio of 0.63. 

In addition, considering the possible difference between spectral fitting procedures in \cite{shen11} and this work (e.g., the use of the iron template, continuum windows, and line fitting method), as a comparison, we also construct a low-redshift control sample using SDSS BOSS spectra and apply our spectral fitting. In order to obtain similar rest-frame spectral coverage to that of our high-redshift quasars with the same continuum windows, we select quasars in a narrow redshift range ($2.0 \le z \le 2.4$) and we also limit the average signal-to-noise ratio in the \civ\ and \mgii\ regions to be higher than seven for good spectral fitting. This sample is much smaller than the sample from \cite{shen11} (our `primary control sample') so we treat this sample as only a secondary control sample. For the selected quasars, we apply the same spectral fitting procedure as that used for our high-redshift quasars and then measure their BH masses and Eddington ratios. We apply the same luminosity matching process as described before. This secondary control sample yields a mean Eddington ratio of 0.74 (median 0.62), consistent with the result from the primary control sample. 

Therefore, although the Eddington ratios of both the high-redshift quasars and the low-redshift control sample vary because of different BH mass estimators or spectral fitting, the high-redshift sample always has a higher mean or median Eddington ratio than the low-redshift control sample.
Since this work focuses only on luminous quasars, the difference in Eddington ratios between our high-redshift quasars and the low-redshift luminosity-matched sample could be explained by the limited BH mass growth in the early Universe.
Under the same luminosity distribution, the low-redshift quasars have BH masses from $9\times 10^{7}$ to $4\times 10^{10}\,M_{\odot}$ (Figure \ref {fig:bh_l_edd}, top). But at $z \gtrsim 6.5$, limited by the available time for BH growth, $10^{10}\,M_{\odot}$ BHs are very rare and thus the majority of these high-redshift quasars have higher Eddington ratios than the low-redshift sample.

\subsection{Iron Templates and Other Potential Uncertainties}
The BH masses and Eddington ratios discussed above are derived based on our spectral fitting using a combined iron template of VW01 and T06, with T06 being used for the \mgii\ region. The \mgii-based BH mass estimator VO09 was originally calibrated based on the VW01 iron template, in which the iron emission underlying the \mgii\ line (2770-2820 \AA) is set to zero, while from theoretical considerations there should be iron emission at this wavelength. Modified or new iron templates have been generated to model the iron emission below the \mgii\ line \citep[e.g., T06;][]{kurk07,derosa11,shen11}. In this work, we use the T06 template for spectral fitting in the \mgii\ region to separate the \mgii\ from the \feii\ emission, which is important for the measurement of spectral properties in the \mgii\ line region (e.g., \mgii\ redshift, \mgii\ FWHM, and \mgii\ flux) and also for the study of other quasar properties (e.g., \civ\ blueshift and \feii/\mgii\ ratio).

The differences between the iron templates VW01 and T06 in the \mgii\ line fitting and BH mass measurements for high-redshift quasars have also been discussed in previous work \citep[e.g., ][]{schindler20, onoue20}. In order to investigate the impact of iron templates on the BH masses of our quasars, we perform the same spectral fitting for all spectra, but using the VW01 iron template, and measuring the corresponding BH masses. The BH masses and Eddington ratios derived from spectral fitting using the VW01 iron template are listed in Table \ref{tab:vw01fitting} in Appendix B.
Compared with the T06-based fitting of the \mgii\ region, the VW01-based fitting yields averaged 1.06 times higher $L_{\rm bol}$ and 1.20 times higher BH masses (0.89 to 2.37 times). For most objects (33/37), the difference in BH masses is $\sim$0.1 dex, much smaller than the systematic uncertainty of the scaling relation (0.55 dex). The mean Eddington ratio derived from VW01 fitting is 1.02 (median 0.88), similar to the determinations based on our combined iron template. Therefore, using the spectral measurements based on the T06 or VW01 templates will not lead to significant changes in the BH masses and Eddington ratios. 
In this paper, we adopt the measurements based on the combined iron template of VW01 and T06 (T06 in the \mgii\ region) as our primary results. 

The commonly used \feii\ templates for quasar spectral fitting are constructed based on the \feii\ emission from low-redshift narrow line Seyfert 1 galaxy, I Zw 1, and are broadened based on the quasar \mgii\ line width when fitting the spectra. The observed \feii\ emission is dependent on a number of factors including the continuum emission of the quasar and the physical conditions of the BLR gas, so it is uncertain whether the template derived from an AGN of much lower luminosity provides an accurate model for the \feii\ emission from these high-redshift quasars. The broadening of the iron template using the \mgii\ line width also effectively assumes that the \feii\ emission originates from the same portion of the BLR as the \mgii\ line. The details of the template and its velocity broadening will therefore lead to additional uncertainties in the measurement of quasar spectral properties. However, in most cases, we find that our modeling of the high-redshift quasar spectra using these iron templates provides good overall fits with small residuals, although lines and continuum parameters may be degenerate.
 
Current determinations of single-epoch virial BH masses of high-redshift quasars are mainly based on scaling relations using the \mgii\ line, which recently have been suggested to potentially include larger intrinsic uncertainties. There is a growing recognition in recent reverberation mapping observations that the quasar ``radius-luminosity'' ($R - L$) relationship is not as tight as was previously assumed, and the correlations between the deviations of the $R-L$ relation and quasar properties could be significant in some cases \citep[e.g.,][]{fonsecaalvarez20}. A possible trend has also been suggested such that objects with high accretion rates have smaller $R_{\rm BLR}$ \citep{du16,du18}, which means that for quasars hosting highly accreting BHs (such as the high-redshift quasars in this paper), the current measurements of BH masses could be overestimated. 
In addition, the \mgii-based scaling relations are calibrated using H$\beta$ measurements. However, recent $R-L$ determinations from the \mgii\ line found an intrinsic scatter of 0.36 dex, significantly larger than that from H$\beta$, implying a broader range of \mgii\ radii than observed for H$\beta$ \citep{homayouni20}.

\section{Rest-frame UV Properties}
\subsection{Broad Emission Line Velocity Shifts}
The velocity shifts of quasar emission lines, especially the velocity differences between high- and low-ionization lines, have already been widely discussed in earlier studies \citep[e.g.,][]{gaskell82, richards02, richards11, derosa14}. The correlations between the line velocity shifts and quasar intrinsic properties (e.g., quasar UV luminosity, line FWHM, or line EW) have also been observed at different redshifts \citep[e.g.,][]{richards02, richards11,schindler20}.
Recent observations of high-redshift quasar samples raise questions about the increase of \civ\ blueshifts at $z \gtrsim 6$ \citep[e.g.,][]{mazzucchelli17, banados18, meyer19, shen19, schindler20}. \cite{meyer19} and \cite{schindler20} report high \civ\ blueshifts in high-redshift quasars, with mean \civ\ blueshifts $\sim$ \hbox{--1500} km\,s$^{-1}$ and $>$ \hbox{--2500} km\,s$^{-1}$ at $z\sim6$ and 6.5, respectively, while quasars at $z \sim 1-4$ have mean \civ\ blueshifts about \hbox{--1000} km\,s$^{-1}$. \cite{shen19}, however, obtain similar \civ\ blueshifts ($\sim$ \hbox{--1000} km\,s$^{-1}$) from both a $z\sim6$ sample and a low-redshift control sample.

As the largest quasar sample at $z > 6.5$, our sample gives new insight into the UV emission line velocity shifts of quasars in the early Universe. 
Since the spectral fitting of \ion{Si}{4} and \ion{C}{3}] is limited by the spectral coverage and low data quality at the edge of the recorded spectra, here we discuss only the \civ\ and \mgii\ line properties. 
We will discuss the \civ\ -- \mgii\ line velocity shift and the shifts of these two lines with respect to the [\cii] redshift below. 
The velocity shift is described in the observer's frame, so a negative value denotes a blueshifted emission line. 

\begin{figure}
\centering 
\epsscale{1.2}
\plotone{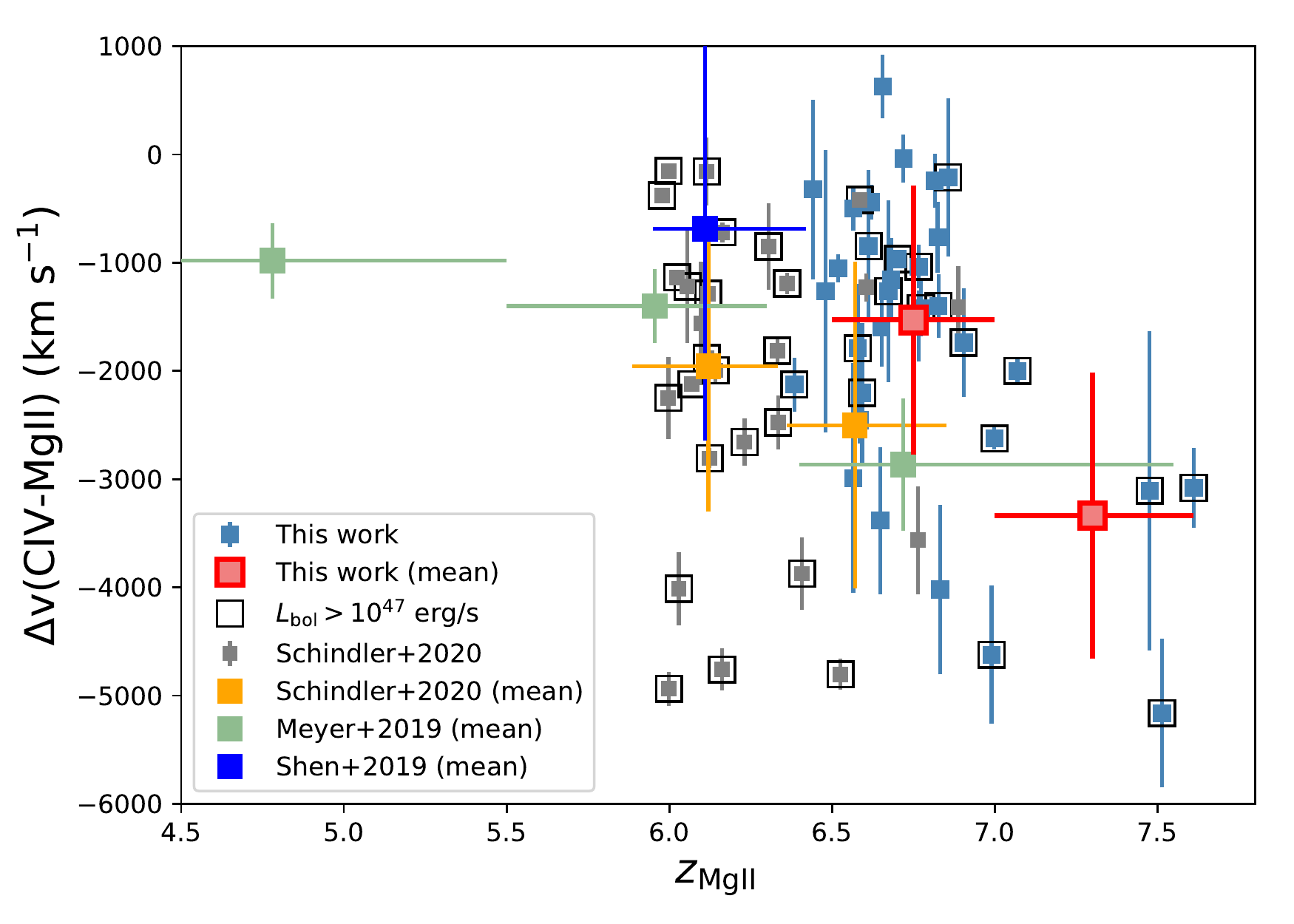} 
\plotone{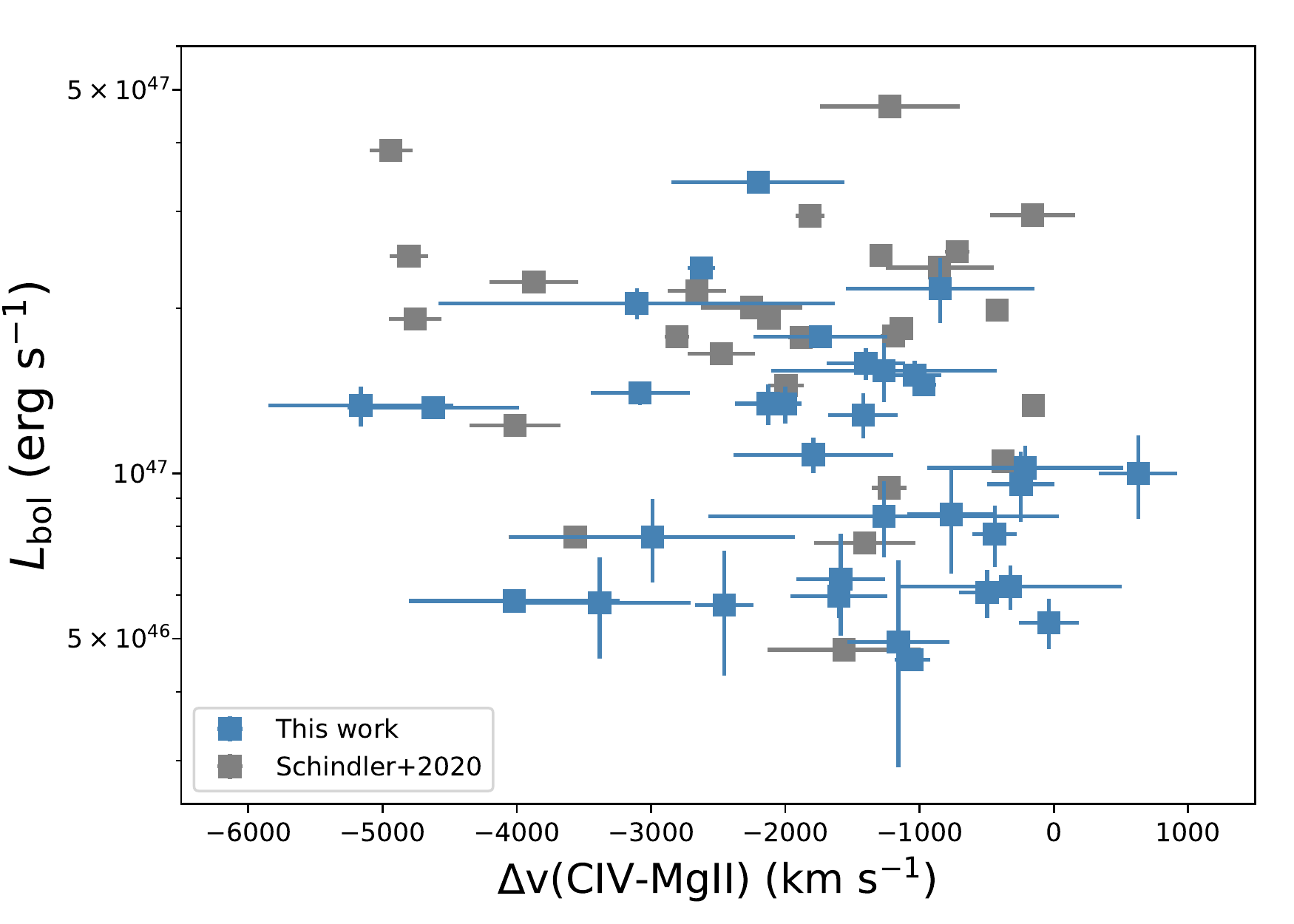} 
\caption{{\bf Top:} The \civ\ -- \mgii\ velocity shifts measured from our sample as a function of the redshift of the \mgii\ line. The light blue squares are obtained from individual quasars and the large red squares with error bars represent the mean values and standard deviations in two redshift bins, $6.5 \le z < 7$ and $z \ge 7$. We compare our measurements with the results from \cite{meyer19} (mean, green squares), \cite{schindler20} (individuals and mean, grey and orange squares), and \cite{shen19} (mean, blue square). Quasars with $L_{\rm bol} > 10^{47}$ erg/s are marked by black open squares, in both our sample and the sample from \cite{schindler20}. Our sample shows a weakly increased \civ\ blueshift at $z = 6.5 -7$ and a significantly higher blueshift at $z > 7$, where only four quasars are measured. The difference between the results from different samples at $z > 6$ could be attributed to small sample size, different luminosity distributions, and different spectral fitting methods. {\bf Bottom:} The \civ\ -- \mgii\ velocity shifts vs.\ bolometric luminosity, including our new sample and the measurements from \cite{schindler20}. No correlation between the \civ\ blueshift and quasar luminosity found in either sample.}
\label{fig:civ}
\end{figure}
 
We measure the redshifts from the line centroids of \civ\ and \mgii\ and then calculate the \civ\ velocity shifts with respect to \mgii, $\Delta v$(\civ\ -- \mgii). Four quasars (J0525--2406,  J0910--0414, J1316+1028, and J1535+1943) do not have this measurement due to the lack of \civ\ fitting. In our sample, the $\Delta v$(\civ\ -- \mgii) values of all 33 quasars span 600 to \hbox{--5100} km\,s$^{-1}$, with a mean of \hbox{--1700} km\,s$^{-1}$. One quasar (J2232+2930) yields a redshifted \civ\ line, caused by a red component of its \civ\ line. We also divide our sample into two redshift bins, $6.5 \le z < 7$ and $z \ge 7$, and calculate the mean at each bin. The mean blueshift in the $z = 6.5 -7$ bin is \hbox{--1500}$\pm$100 km\,s$^{-1}$ and \hbox{--3300}$\pm$400 km\,s$^{-1}$ in the higher redshift bin. As a comparison, we also calculate the mean blueshifts using the line redshifts derived from line peaks instead of from line centroids, and find them to be \hbox{--1700}$\pm$100 km\,s$^{-1}$ in the $z = 6.5 -7$ bin and \hbox{--3100}$\pm$200 km\,s$^{-1}$ at $z\ge7$. Therefore, although the measurements from the line centroid and line peak might be different for a single object, the statistical results of the quasar sample are quite similar. The velocity shifts used for further discussion are all derived from the line centroids.

We then compare our measurements with the results from other investigations at similar redshifts, as shown in Figure \ref{fig:civ}. In general, at $z \sim 6.5-7$ our quasars have \civ\ blueshifts in a similar range to that of the sample from \cite{schindler20}. The four luminous quasars at $z>7$ are all located at the high \civ\ blueshift end (J0038--1527 and J0252--0503 have \mgii-based redshifts below seven so they are not included in this bin). The mean blueshift in the $z = 6.5 - 7$ bin, regardless of how the redshifts are measured, is significantly smaller than the results from both \cite{schindler20} (\hbox{--2501} km\,s$^{-1}$ at $z=6.57$) and \cite{meyer19} (\hbox{--2867} km\,s$^{-1}$ at $z=6.72$) at similar redshifts.  
Note that the results in \cite{schindler20} and \cite{meyer19} are all derived from line peaks, but as mentioned above, for our sample, the mean values from the line centroid and line peak are very similar. One reason for our smaller mean \civ\ blueshift could be the contributions from relatively faint quasars. Our sample includes three times more quasars with luminosities $< 10^{47}$ erg s$^{-1}$ than the sample in \cite{schindler20}, and most of these less luminous quasars have smaller blueshifts than the mean (Figure \ref{fig:civ}). In the $z = 6.5 -7$ bin, these less luminous quasars have a mean of \hbox{--1400} km\,s$^{-1}$, while the mean of the other quasars is --1700 km\,s$^{-1}$, although they are consistent within the uncertainties, and we do see a few faint quasars with large \civ\ blueshifts, as shown in Figure \ref{fig:civ}. 

In addition, the small sample statistics and different spectral fitting methods will also bias the measurements using these high-redshift samples. Our luminous subsample ($> 10^{47}$ erg s$^{-1}$) also has a smaller mean blueshift than the results from the other two investigations at $z \sim 6.5-7$. \cite{schindler20} use nine quasars at $6.36 < z < 6.85$, and \cite{meyer19} include eleven quasars at $6.4 < z < 7.6$. Our sample has 26 quasars in this redshift range but it is still a small sample for estimating a representative value for the entire quasar population. 
Similarly, we can also see a discrepancy at $z \sim 6$. \cite{meyer19} find increased \civ\ blueshift at $z \sim 6$ compared to lower redshift measurements. \cite{schindler20} obtain an even higher blueshift at similar redshift, while \cite{shen19} obtain a mean \civ\ blueshift similar to low-redshift samples and suggest no redshift evolution. 
In the $z>7$ bin, our quasars have a significantly larger mean \civ\ blueshift, but only four quasars are included. 
In particular, the only three $z\sim 7.5$ quasars all have large \civ\ blueshifts. The sample size is too small to represent quasars at $z>7$. 

Therefore, based on our results and the comparisons with other samples, we conclude that from current observations there is a potential increase of \civ\ blueshift toward higher redshift at $z > 6$ but the observed trend of redshift evolution varies among different samples. Our sample shows a weaker redshift evolution than the significant evolution suggested by \cite{meyer19} and \cite{schindler20}. An increase of \civ\ blueshift at high redshift is possible, but the exact evolution is still not clear considering the small sample size, especially at $z \ge 7$, and the difference between spectral fitting methods. In addition, from this sample we do not find any correlations between the \civ\ blueshift and quasar luminosity or Eddington ratio, probably because we are still looking at a narrow $L$ or $\lambda_{\rm Edd}$ range, although the less luminous sample has a relatively small mean of the \civ\ blueshift, as described above (also Figure \ref{fig:civ}). 
The physical reason for the potential increase of \civ\ blueshift is also unclear.
High \civ\ blueshifts have been commonly considered to be associated with strong BLR outflows or winds, which may explain the possible redshift evolution of the \civ\ blueshift as the result of stronger outflows in early quasars. On the other hand, such outflows or winds are also suggested as the reason for strong BAL features. In some samples, it has been found that BAL quasars have somewhat larger \ion{C}{3}] blueshifts \citep{richards11}.
In our sample, we do not find a correlation between high \civ\ blueshifts and strong BAL quasars (see Section 6.2).

\begin{figure}
\centering 
\epsscale{1.2}
\plotone{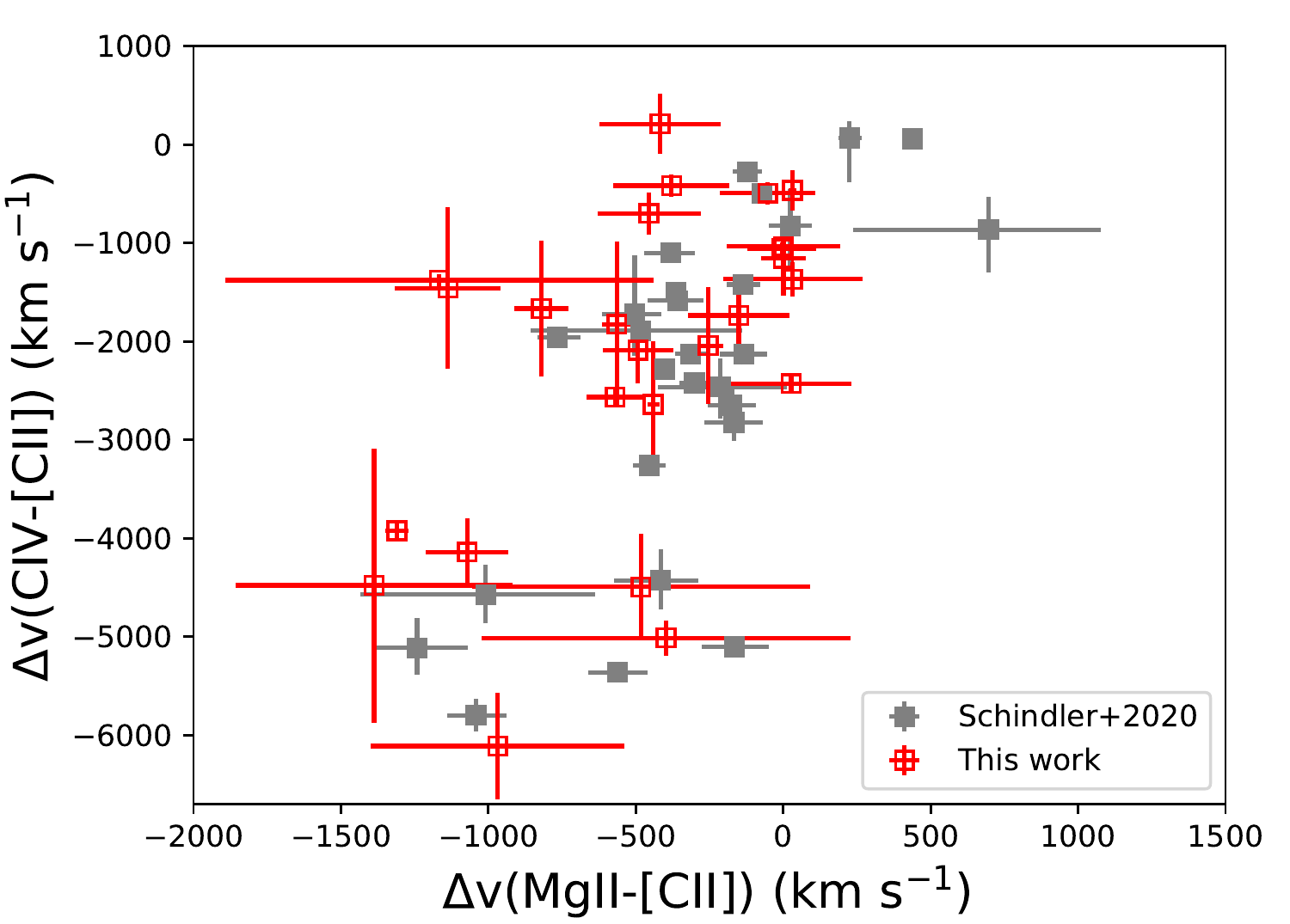} 
\caption{The relation between the \civ\ blueshift and \mgii\ blueshift with respect to the [\cii] redshift. We plot our results (red) and the sample from \cite{schindler20} (grey). Both samples show that there is a correlation between the \civ\ and \mgii\ blueshifts, possibly indicating related physical origins of the velocity shifts of these two lines. There is also a trend that these quasars are distributed in two populations, one with much stronger blueshifts of both the \civ\ and \mgii\ lines than the other one.}
\label{fig:civmgii}
\end{figure}

We also calculate the velocity shifts of the \civ\ and \mgii\ emission lines with respect to the [\cii] line, which represents the systemic redshift of the quasar, and investigate the possible correlation between the velocity shifts of the two lines. 
In our sample, there are 27 quasars that have [\cii]-based redshift measurements and we calculate the $\Delta v$(\civ\ -- [\cii]) and $\Delta v$(\mgii\ -- [\cii]) for each quasar, as plotted in Figure \ref{fig:civmgii}.
Our sample has a mean $\Delta v$(\civ\ -- [\cii]) of \hbox{--2200} km\,s$^{-1}$ and mean $\Delta v$(\mgii\ -- [\cii]) of \hbox{--500} km\,s$^{-1}$. Our results show a correlation between the \civ\ blueshift and \mgii\ blueshift relative to the [\cii] line, which is also reported in \cite{schindler20}. This correlation suggests a potential relation between the physical origins of the blueshifts, although \civ\ is a high ionization line and \mgii\ is a low ionization line, and they are supposed to originate from different locations with different physical conditions in the broad line region. In Figure \ref{fig:civmgii}, there is also a trend that these quasars are in two populations, and quasars in one subsample have much larger velocity shifts.
With limited data available we can only speculate on the nature of these two emerging populations. Potentially the separate distributions might indicate different origins of the line blueshifts, may be tied to different geometrical structure of the BLR, or may be dependent on the orientation to the observer's line-of-sight.

\subsection{Composite Spectrum for $z > 6.5$ Quasars}
\begin{figure*}
\centering 
\epsscale{1.15}
\plotone{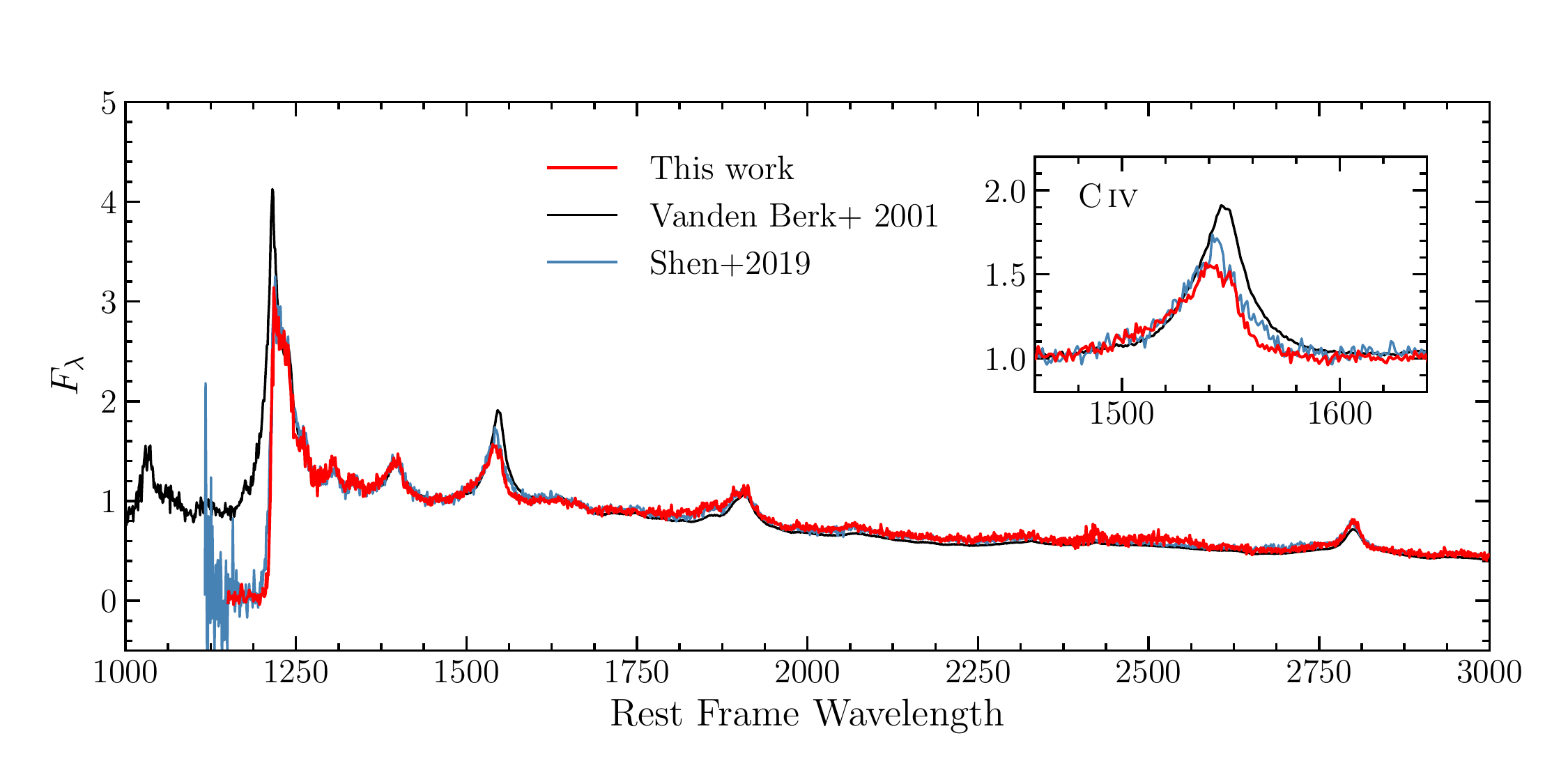} 
\caption{Quasar composite spectrum (red solid line) from our sample compared with the low-redshift composite from \cite{vandenberk01} (black line) and the $z\sim6$ quasar composite from \cite{shen19} (blue line). All three composite spectra have been normalized to the continuum flux at 1450 \AA. The inset shows the \civ\ line region, with a more pronounced blueshift of \civ\ in our composite, compared to both the $z \sim 6$ and low-redshift composites. The low-redshift sample includes a higher fraction of low luminosity quasars than our sample, while the $z\sim6$ quasar sample has a similar luminosity range to our sample.}
\label{fig:composite}
\end{figure*}

In this Section, we present a $z > 6.5$ quasar composite based on our quasar sample and compare it with composite spectra for quasars at different redshifts to discuss the possible redshift evolution of average quasar UV spectral properties. We also include NIR data of quasars in \cite{schindler20} for a larger sample. We choose all 31 $z>6.5$ quasars (excluding J0525--2406; see below for details) in our sample and 7 quasars from \cite{schindler20}, after excluding 2 overlapping quasars between the two samples. 

We generate the median composite spectrum following \cite{vandenberk01} and compare our result with the composite spectra created for SDSS low-redshift quasars in \cite{vandenberk01} and for a sample of $z\sim6$ quasars in \cite{shen19}.
Given that the low-redshift and $z\sim 6$ composite spectra are all based on the redshifts derived from UV/optical emission lines, here we use the \mgii-based redshift instead of the [\cii] redshift. Otherwise, we would see a blueshift of the \mgii\ line compared to the other two composite spectra.
The quasar spectra are all normalized to the rest-frame 1600--1610 \AA\ flux where there are no strong broad lines or iron emission. We use 1600 \AA\ instead of 1450 \AA\ since for BAL quasars, the region around 1450 \AA\ is affected by strong absorption features. 
We exclude the quasar J0525--2406 due to the lack of data at rest-frame $< 1900$ \AA. 
For BAL quasars, we mask all strong absorption troughs. We also mask all strong absorption features visually identified in these spectra, mainly from intervening metal absorbers.
We also compare the composite spectrum including and excluding all BAL quasars and do not find significant differences. The composite plotted in Figure \ref{fig:composite} is the one including BAL quasars.

As shown in Figure \ref{fig:composite}, our composite covers the wavelength range from rest-frame 1150 to 3000 \AA. The data are also provided in Table \ref{tab:composite} and available online\footnote{\url {https://jinyiyang.github.io/composite.html}}.
The composite spectrum has relatively low quality at rest-frame 1750--1850 \AA\ and 2300--2500 \AA, and in particular at 2400--2450 \AA, because of the strong telluric absorption between the observed frame wavelengths of $\sim$ 13500--14200 \AA\ and 18000--19500 \AA. Our sample has a relatively narrow redshift range so only a few spectra could be used to fill these two gaps. The wavelength range of rest-frame 2400--2450 \AA\ is covered by only five spectra. At wavelengths other than these two regions, our composite is based on 17 to 38 quasar spectra at each pixel.

\begin{deluxetable}{l c c}
\tablecaption{$z>6.5$ Quasar Composite Spectrum}
\setlength{\tabcolsep}{15pt}
\tabletypesize{\scriptsize}
\tablewidth{0pt}
\tablehead{
\colhead{Wavelength (\AA)} &
\colhead{$f_{\rm \lambda}$} &
\colhead{N$_{\rm QSO}$}
}
\startdata
1150.5 & --0.019 & 17\\
1151.5 & 0.095 & 17\\
1152.5 & 0.029 & 17\\
... & & \\
2080.5 & 0.716 & 38\\
2081.5 & 0.744 & 37\\
... & & \\
2997.5 & 0.439 & 26\\
2998.5 & 0.461 & 25\\
\enddata
\label{tab:composite}
\tablenotetext{}{{\bf Note.} The median composite spectrum for our quasar sample. Wavelengths are in the rest frame and in units of \AA. Flux density units are arbitrary. The last column shows the number of quasar spectra contributing to the composite spectrum at each pixel. This is only a portion of the full table, and the entire table data is available online.}
\end{deluxetable}

The composite spectra from \cite{vandenberk01} and \cite{shen19} are both median composites; the former was created using over 2200 SDSS quasars at $0.044 \le z \le 4.789$, and the latter was based on 50 quasars at $5.71 < z < 6.42$. The low-redshift quasar sample mostly has luminosity lower than our quasars and the $z \sim 6$ sample has a luminosity range similar to that of our sample.
Compared with these two composite spectra, our composite for $z>6.5$ quasars has very similar line strengths in most broad emission lines. Our quasars have a weaker Ly$\alpha$ line, significantly weaker than SDSS low-redshift quasars and slightly weaker than the $z\sim 6$ sample, because of the increasing absorption from the intergalactic medium toward higher redshift.
Both our composite and the $z \sim 6$ composite exhibit an obvious blueshift of the \civ\ line relative to the SDSS low-redshift composite, mainly caused by the difference in luminosity, which has been suggested by the comparisons between composite spectra for the $z \sim 6$ sample and a low-redshift luminosity-matched sample in \cite{shen19}. The \civ\ line in our composite is also blueshifted relative to the $z \sim 6$ composite, which is consistent with the results discussed in Section 5.1, indicating a potential redshift evolution of \civ\ blueshift. 
All three of these composite spectra have consistent continuum slopes, indicating no strong evolution of the continuum, although the median composite is more suitable for studying the relative fluxes of emission lines \citep{vandenberk01}.

\subsection{\feii/\mgii\ up to $z = 7.6$}
Observations of $z \gtrsim 6$ quasars have found that these early SMBHs are accompanied by intense star formation and high metallicity in their environments. 
Photoionization models show that quasar emission-line ratios provide estimates of metallicity in the broad-line region \citep[e.g.,][]{hamann93,hamann02, nagao06}. At high redshift, UV emission line ratios, such as \nv/\civ, \nv/\heii, and \feii/\mgii, have been used to characterize the BLR metallicity of distant quasars \citep[e.g.,][]{hamann99,jiang07,derosa11,onoue20,schindler20}. In addition,  Fe/$\alpha$ is expected to be a useful probe of the gas chemical enrichment history in these early quasar environments. 
In the local Universe, Fe is mainly produced by Type Ia supernovae, while the production of $\alpha$-elements like Mg and O is dominated by Type II, Ib and Ic supernovae. Therefore, appreciable Fe enrichment is expected to have a $\sim$ 1 Gyr delay after $\alpha$-element enrichment \citep[e.g.,][]{greggio83}.
This has led to the expectation that there might be a decrease in \feii/\mgii\ with increasing redshift in quasars at redshifts above 6, corresponding to 0.92 Gyr after the Big Bang.

\begin{figure*}
\centering 
\epsscale{1.2}
\plotone{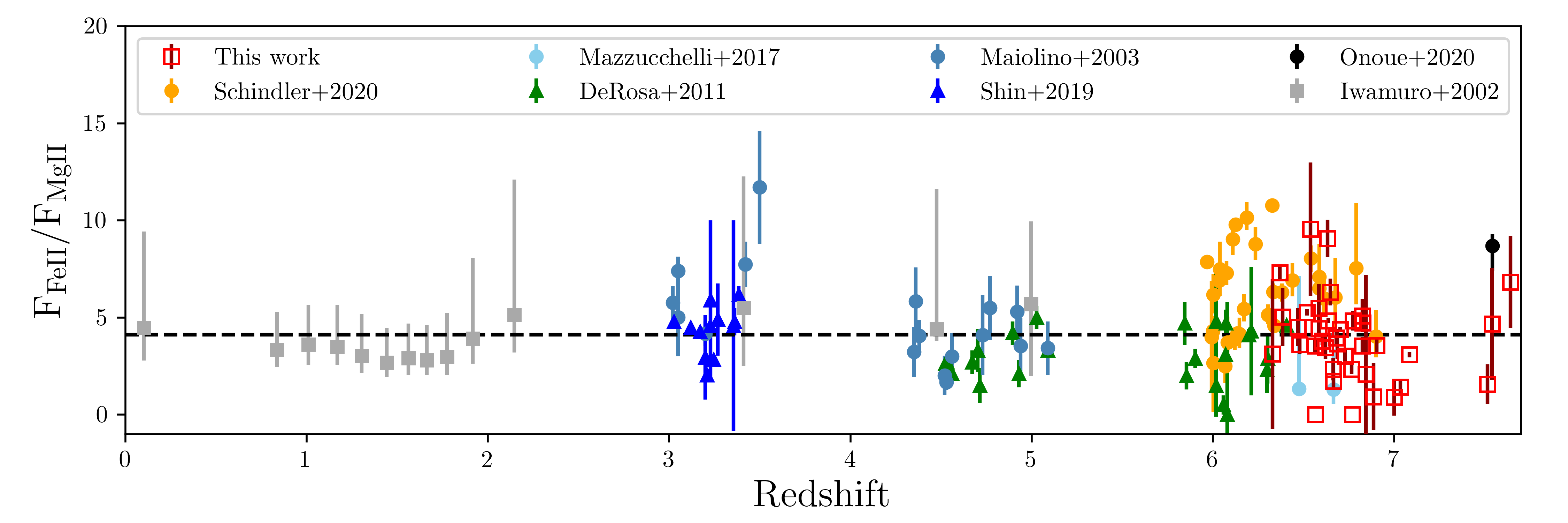} 
\caption{The \feii/\mgii\ flux ratio of quasars in our sample, compared with measurements at different redshifts in the literature \citep{iwamuro02, maiolino03, derosa11, mazzucchelli17, shin19, schindler20, onoue20}. The dashed line represents the mean (4.1) of our sample.
These measurements suggest no significant redshift evolution of \feii/\mgii\ up to redshift 7.6, only 0.67 Gyr after the Big Bang, although the comparison can be affected by the uncertainties caused by different spectral fitting procedures (e.g., different iron templates). The results from \cite{shin19}, \cite{schindler20}, and \cite{onoue20} are based on the iron template from T06, while \cite{maiolino03}, \cite{derosa11}, and \cite{mazzucchelli17} apply the iron template from VW01. \citet{iwamuro02} use their own iron template.}
\label{fig:feii_mgii}
\end{figure*}

However, observations of quasars up to $z \sim 7.5$ have not shown any evidence of such evolution. Using our sample, including more $z > 6.5$ quasars, we measure the \feii/\mgii\ flux ratio to test its redshift evolution. The \feii\ flux is derived by integrating the best-fit \feii\ component over the rest-frame wavelength range 2200 to 3090 \AA. We then compare our results with such measurements from other investigations from $z<1$ to $z=7.5$ \citep{iwamuro02,maiolino03, derosa11,mazzucchelli17, shin19, onoue20,schindler20}, as displayed in Figure \ref{fig:feii_mgii}. The quasars in our sample have \feii/\mgii\ values comparable to those of quasars at similar or lower redshifts, demonstrating that there is no significant evolution up to redshift 7.6, although the measurements from different investigations may have different systematic uncertainties. Our sample has a mean \feii/\mgii\ ratio of 4.1 (median value of 3.8), which agrees with most samples at all redshifts. 

The absence of strong redshift evolution of \feii/\mgii\ from the current epoch up to redshift 7.6 (0.67 Gyr after the Big Bang) could be explained by scenarios of shorter timescales for SNe Ia or different origins of Fe \citep[e.g.,][]{jiang07, onoue20}.
\cite{rodney14} find that the fraction of prompt SNe Ia which explode within 500 Myr could be as high as $\sim$50\%.
It has also been suggested that the timescale of maximum chemical enrichment from SNe Ia is a strong function of star formation history in galaxies and could be as short as $\sim 0.3$ Gyr in specific galaxy environments \citep{matteucci01}. If a short timescale for SNe Ia is the correct explanation, star formation in these high-redshift quasar host galaxies needs to occur very early. 
Different origins of iron enrichment, such as Population III stars, could also explain this lack of evolution. It has been suggested that these massive stars could produce a large amount of Fe within a few Myr \citep{heger02}.

We also note that the spectra of two quasars, J0218+0007 and J2338+2143, have almost zero \feii\ flux measured from spectral fitting. Both of them have relatively lower S/N in the continuum, so the results could be lower limits. Quasars at $z > 6$ with very low \feii/\mgii\ have also been found in previous studies \citep[e.g.,][]{mazzucchelli17}, suggesting large scatter of iron abundance in these high-redshift quasar BLRs, and may indicate that we are witnessing ongoing iron enrichment in these BLRs. However, these results are affected by a number of factors, including the spectral quality, different spectral fitting methods (e.g., different iron templates, the choice of continuum windows, and the \mgii\ line fitting procedure, as discussed in detail by \citealt{schindler20} and \citealt{onoue20}), and also the possible Eddington ratio dependence of \feii/\mgii\ \citep[e.g.,][]{sameshima17}. These factors could result in different systematic uncertainties in different samples.

\section{Discussion}
\subsection{Early SMBH Growth}
The recent discoveries of SMBHs with $\gtrsim 10^9\,M_{\odot} $ at $z > 6$, in particular at $z> 7$, have already posed significant challenges to BH formation theories \citep[e.g,][]{mortlock11, wu15, banados18, yang20a, wang21b}. The existence of these SMBHs requires either very massive initial seeds or super/hyper-Eddington accretion \citep[e.g.,][]{volonteri12, pacucci17, inayoshi19}.  
As described above (Section 3), our quasars yield a sample of 37 SMBHs at $z = 6.3-7.6$ with masses in the range $2.6 \times10^{8} -3.6\times10^{9}\,M_{\odot}$, and they have a mean Eddington ratio of 1.1, with a peak at $\lambda \sim 0.8$. With these measurements for a large sample of SMBHs at $z \gtrsim6.5$, we are able to revisit the BH growth scenario and seek new constraints.

\begin{figure}
\centering 
\epsscale{1.15}
\plotone{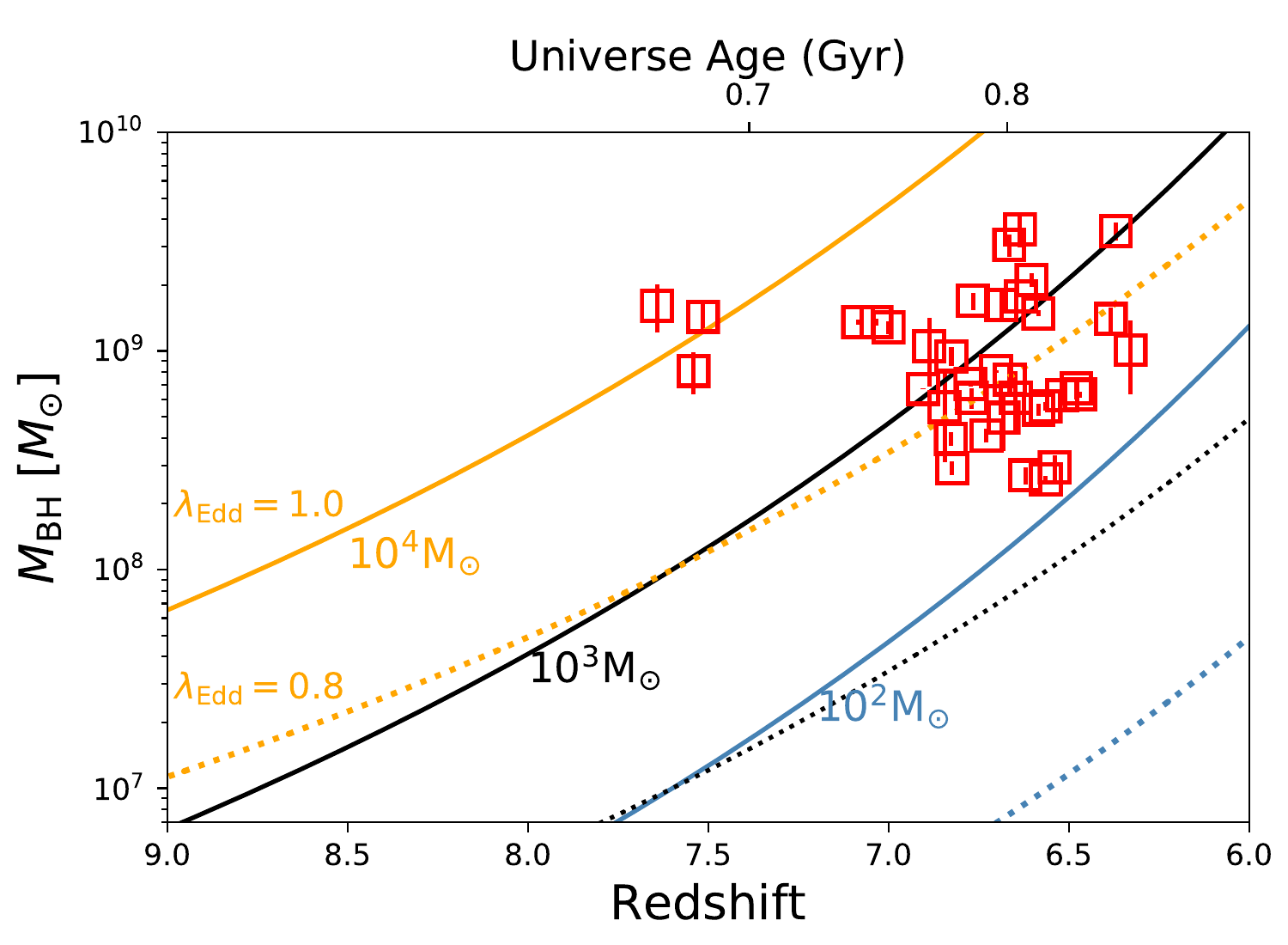} 
\caption{BH mass measurements (red open squares) obtained from our quasar sample compared with the BH growth tracks with different seed BH masses. The three solid curves represent the BH growth tracks with seed BHs masses of $10^{2}\,M_{\odot}$ (blue), $10^{3}\,M_{\odot}$ (black), and $10^{4}\,M_{\odot}$ (orange), respectively, assuming Eddington accretion since $z=30$. The three dotted lines are the BH growth tracks with constant Eddington ratio $\lambda_{\rm Edd}=0.8$, which is the peak of the Eddington ratio distribution from our sample. All these tracks are based on the assumption of a radiative efficiency of 0.1. With Eddington accretion, most of the quasars in our sample require massive seed BHs with masses $\gtrsim 10^{3}\,M_{\odot}$ at $z=30$, and the three $z=7.5$ quasars require $\gtrsim 10^{4}\,M_{\odot}$ seed BHs. A later starting time for BH growth, lower accretion rate, or higher radiative efficiency will result in a requirement of even more massive seed BHs. With $\lambda_{\rm Edd}=0.8$, most of the quasars will need $\gtrsim 10^{4}\,M_{\odot}$ BH seeds at $z=30$.}
\label{fig:bhgrowth}
\end{figure}

BH growth can be modeled according to $M_{\rm BH} = M_{\rm seed} \rm exp[ \lambda_{Edd}(1-\epsilon)t/4.5\times10^{8} \epsilon]$, where $\epsilon$ is the radiative efficiency and a typical value of 0.1 has been suggested for $z \gtrsim 5.7$ quasars \citep[e.g.,][]{trakhtenbrot17}. We first assume that these BHs grew at the Eddington limit across the entire time since BH seeding (that is, assuming an accretion duty cycle of unity) with a radiative efficiency of 0.1. With these assumptions, we obtain BH growth tracks since $z=30$ for seeds with different BH masses, as shown in Figure \ref{fig:bhgrowth}. With Eddington accretion, most of the quasars in our sample require seed BH masses $\gtrsim 10^{3}\,M_{\odot}$ at $z=30$ (or $>10^{4}\,M_{\odot}$ at $z = 15$), and the three $z=7.5$ quasars require $\gtrsim 10^{4}\,M_{\odot}$ seed BHs at $z=30$. Only a few quasars allow less massive seed BHs down to a few hundred solar masses. We also consider a lower Eddington ratio for these quasars, since the Eddington ratio distribution obtained from our sample has a peak at $\lambda_{\rm Edd}\sim0.8$. 
Assuming an Eddington ratio of 0.8, most of our quasars would require at least $10^{4}\,M_{\odot}$ seeds to grow to their measured masses starting from $z=30$. Note that we assume a start time of BH growth at $z = 30$, while only stellar-mass BHs have been suggested to form at such high redshifts \citep{inayoshi19}. A later starting time for BH growth or a higher radiative efficiency would therefore require more massive seed BHs. 

Based on the discussion above, in most cases, the SMBHs in these luminous quasars need to grow from BH seeds more massive than a few hundred solar masses (i.e., the masses consistent with Pop III stellar remnants). 
In particular, the results from the three $z=7.5$ quasars are more consistent with massive seed BH models like direct collapse of gas, which however are suggested to occur only in  rare and special environments \citep[e.g.,][]{haiman13}.
It is considered difficult for $100\,M_{\odot}$ seeds to grow to the SMBH masses observed for our sample over such a short duration, although this can not be ruled out due to the unclear BH accretion history. With a seed BH of $\sim 100\,M_{\odot}$, it would take 0.8 Gyr to grow to a $1\times 10^{9}\,M_{\odot}$ BH assuming Eddington accretion and a radiative efficiency of 0.1, which is impossible for all $z > 6$ quasars.
It is also thought to be highly unlikely that BHs can have sustained growth for the full 0.8 Gyr \citep[e.g.,][]{davies19, eilers20}.
Super/hyper-Eddington accretion has been proposed to occur under specific conditions \citep[e.g.,][]{inayoshi16} and it has been suggested that such processes may dominate BH growth until $z \sim 10$ \citep[e.g.,][]{pezzulli16}. However, the rate of occurrence of extreme high accretion rates or how to maintain a long-term high accretion rate are still open questions. 

To date, BH masses have been measured for about 100 $z > 6$ quasars using the \mgii\ line. Only one has been discovered with a BH mass exceeding $10^{10}\,M_{\odot}$ \citep{wu15}.
The currently known high-redshift quasars are selected from flux-limited surveys, typically down to a luminosity of $5\times10^{46}$ erg s$^{-1}$ for a $z\sim6.5$ quasar. Most quasars are selected based on photometric data. There is no obvious reason why this selection method should result in significant incompleteness at the massive end of the BH mass distribution in unobscured quasars. Therefore, if quasars with $10^{10}\,M_{\odot}$ BHs exist at this redshift, they should mostly have Eddington ratios $< 0.05$ or significant obscuration. 

\subsection{BAL Fraction}
From our sample, we calculate the balnicity index BI \citep{weymann91} by
\begin{equation}
{\rm BI} = \int_{v_{\rm min}}^{v_{\rm max}}\left ( 1 - \frac{f(v)}{0.9} \right )C dv.
\end{equation}
where $f(v)$ is the normalized spectrum, and $C$ is set to 1 only when $f(v)$ is continuously smaller than 0.9 for more than 2000 km s$^{-1}$, otherwise it is set to 0.0. The value of $v_{\rm min}$ is set to 0.
We identify nine quasars with strong broad absorption features, including J0038--1527, J0246--5219, J0313--1806, J0439+1634, J0706+2921, J0839+3900, J0910--0414, J0923+0402, and J1316+1028, as indicated in Table \ref{tab:fitting}. This results in a BAL fraction of 24\% of this sample. The quasar J1316+1028 shows an obvious BAL feature in its optical spectrum \citep[see Figure 2 in][]{wang19}, while its NIR spectrum does not show $>$ 2000 km\,s$^{-1}$ continuous absorption (with normalized flux density $<$ 0.9). The quasars J0803+3138 and J0837+4929 also have absorption troughs, but the widths of their troughs are close to the limit of 2000 km\,s$^{-1}$, and the BI calculations give results that are highly dependent on spectral fitting details. Also, there are no obvious BAL features in their optical spectra. We thus do not include these two quasars when we estimate the BAL fraction. J0525--2406 only has NIR spectral coverage at $> 14000$ \AA\, and no strong features are present within this wavelength range or in its optical spectrum ($< 1 \mu$m). Considering the limited S/N and spectral coverage at rest-frame wavelengths 1400--1500 \AA\ in some of the spectra, the BAL fraction reported here could be a lower limit.

A BAL fraction of 24\% in this sample is higher than the results from lower redshift samples, 16\% at $z \sim 6$ \citep{shen19} and 15\% from the SDSS DR5 low-redshift sample \citep{gibson09}.
The quasar sample in \cite{schindler20} has comparable size to our sample and is in a redshift range ($5.78\le z \le 7.54$) closer to our sample than others. A BAL fraction of 13\% (5/38) has been claimed in their sample through visual classification. If we simply combine the two samples, we would obtain a BAL  fraction of 19\% (14/72, after removing 3 overlapping quasars between the two samples). If we take into account only $z>6.5$ quasars, with 32 quasars (8 BAL quasars) from our sample and 7 quasars (1 BAL) from \cite{schindler20}, we derive a fraction of 23\% (9/39) in the combined sample. 
The high fraction of strong BALs at $z> 6.5$ from our sample and the combined sample could be caused by a higher intrinsic BAL fraction at high redshift or a selection bias.
The current high-redshift quasar samples are still small, so they may be subject to biases in the quasar selection.

We now compare the spectral properties of these BAL quasars with the non-BAL quasars in our sample. The nine BAL quasars have a median continuum slope of $-1.0$, slightly redder than non-BAL quasars with a median slope of $-1.3$. In addition, these BAL quasars have red $J-W1$ colors with a median value of 2.5, while the median color of the non-BAL quasars is 2.1. So the BAL quasars in our sample do have redder colors than the non-BAL quasars, although this might not represent the intrinsic reddening of the BAL population at this redshift due to the small sample size. 
BAL features are thought to be associated with powerful outflows or disk winds, which are also commonly considered as the origin of high blueshifts of the \civ\ lines. Thus we test the potential correlation between the BAL features and \civ\ blueshifts, but no difference is found between BAL and non-BAL quasars. The mean \civ\ blueshift of BAL quasars is --1700 km\,s$^{-1}$, similar to the mean of non-BAL quasars and of the entire sample. Note that the \civ\ fitting for BAL quasars could be affected by the absorption troughs.
Overall, the high BAL fraction (24\%) in this sample, compared to the fractions at lower redshift (16\% at $z\sim6$; 15\% at lower redshift), potentially indicates a high probability of strong outflows or winds, which may also be an explanation of the higher \civ\ blueshift observed in high-redshift quasars.

\section{Summary}
We report our studies of quasar BH mass and UV spectral properties using a new NIR spectral dataset of 37 luminous quasars at $6.3 < z \le 7.64$, with 32 quasars at redshift above 6.5, forming the largest quasar NIR spectral sample at this redshift to date. 
The NIR spectroscopy was obtained using the Keck/NIRES, Gemini/GNIRS and F2, VLT/XShooter, and Magellan/FIRE instruments. These data allow us to model quasar rest-frame UV spectra, statistically characterize quasar UV spectral properties, and study quasar BH behavior. We summarize the main results below.

\begin{itemize}
\item We measure the BH masses of these 37 quasars using uniform spectral fitting procedures and BH mass estimators. These objects have BH masses in the  range $(0.3-3.6)\times10^{9}\,M_{\odot}$. Assuming Eddington-limited accretion, they require massive seed black holes with masses $\gtrsim10^{3-4}\,M_{\odot}$ at $z=30$.

\item Luminous quasars in this sample are found to be accreting close to the Eddington limit at the observed epoch. The Eddington ratio distribution has a mean of 1.08 and a median of 0.85,  with a peak at 0.8, significantly higher than the Eddington ratios from a low-redshift luminosity-matched quasar sample. 

\item The difference between \civ\ and \mgii\ redshifts suggests a large blueshift of the \civ\ lines, yielding a mean of --1500 km\,s$^{-1}$ at $z = 6.5-7$ and --3300 km\,s$^{-1}$ at $z\ge7$. Compared with $z \sim 6$ samples of similar luminosity, our results show an increase of \civ\ blueshift at $z> 6$, but the evolution is weaker than previously reported. The correlation between \civ\ and \mgii\ line velocity shifts with respect to the [\cii] line redshift potentially indicates associated origins of velocity shifts of these two lines.

\item We create a $z > 6.5$ quasar composite spectrum using 38 $z>6.5$ quasars and compare it with the composite spectra for low-redshift SDSS quasars and $z\sim6$ quasars. No significant redshift evolution is found for either broad UV emission lines or quasar continuum slope, except for the \civ\ line which shows a blueshift relative to both low-redshift and $z\sim6$ composites.

\item We measure the \feii\ to \mgii\ flux ratio for quasars in this sample and compare the results with measurements at different redshifts. No redshift evolution of \feii/\mgii\ is found up to redshift 7.6, suggesting that the metal abundances in the BLRs of these quasars are similar to those observed at lower redshifts. 

\item We identify strong BAL quasars and find a BAL fraction of 24\%, higher than the fractions in lower redshift samples. The high BAL fraction could be due to evolution of intrinsic properties (e.g., stronger outflows or winds) of these quasars or due to selection effects in high-redshift quasar surveys.

\end{itemize}

In subsequent work, we will combine data from this NIR sample with our optical spectral dataset \citep{yang20b} as well as a submm dataset from ALMA, NOEMA, and JCMT to further investigate the quasar proximity zones, star formation in the host galaxies, BH-host co-evolution, and the potential correlations among these properties.

\acknowledgments
J. Yang, X. Fan, and M. Yue acknowledge support from US NSF grants AST 15-15115, AST 19-08284 and NASA ADAP Grant NNX17AF28G.
F. Wang and ACE acknowledge support by NASA through the NASA Hubble Fellowship grant \#HST-HF2-51448.001-A and $\#$HF2-51434 awarded by the Space Telescope Science Institute, which is operated by the Association of Universities for Research in Astronomy, Incorporated, under NASA contract NAS5-26555.
Research at UC Irvine was supported by NSF grant AST-1907290. Some of the data presented herein were obtained using the UCI Remote Observing Facility, made possible by a generous gift from John and Ruth Ann Evans.
E.P. Farina acknowledge support from the ERC Advanced Grant 740246 (Cosmic Gas).
L. J. and X.-B. Wu acknowledge support by the National Key R\&D Program of China (2016YFA0400703) and the National Science Foundation of China (11721303, 11890693).
We acknowledge Dale Mudd for his help with the Keck/NIRES observations.

Some of the data presented in this paper were obtained at the W.M. Keck Observatory, which is operated as a scientific partnership among the California Institute of Technology, the University of California and the National Aeronautics and Space Administration.
The Observatory was made possible by the generous financial support of the W. M. Keck Foundation.
The authors wish to recognize and acknowledge the very significant cultural role and reverence that the summit of Maunakea has always had within the indigenous Hawaiian community. We are most fortunate to have the opportunity to conduct observations from this mountain.
This work is based in part on observations made with ESO telescopes at the La Silla Paranal Observatory under program ID 0103.A-0423(A).
This paper also uses data based on observations obtained at the Gemini Observatory, which is operated by the Association of Universities for Research in Astronomy, Inc., under a cooperative agreement with the NSF on behalf of the Gemini partnership: the National Science Foundation (United States), the National Research Council (Canada), CONICYT (Chile), Ministerio de Ciencia, Tecnolog\'{i}a e Innovaci\'{o}n Productiva (Argentina), and Minist\'{e}rio da Ci\^{e}ncia, Tecnologia e Inova\c{c}\~{a}o (Brazil).
This paper includes data gathered with the 6.5 meter Magellan Telescopes located at Las Campanas Observatory, Chile. We acknowledge the use of the MMT and UKIRT telescopes.
We acknowledge the use of the PypeIt data reduction package.

%

\vspace{5mm}
\facilities{Keck(NIRES), Gemini(GNIRS,F2), VLT(X-Shooter), Magellan(FIRE), MMT(MMIRS), UKIRT(WFCAM)}


\software{PypeIt}



\appendix
\section{Quasar Spectra Fitting}
Here we plot the spectra and best-fits of all quasars in our sample, except for quasar J0319--1008, which has been shown in Figure \ref{fig:fitting}.

\begin{figure*}
\centering 
\epsscale{1.2}
\plotone{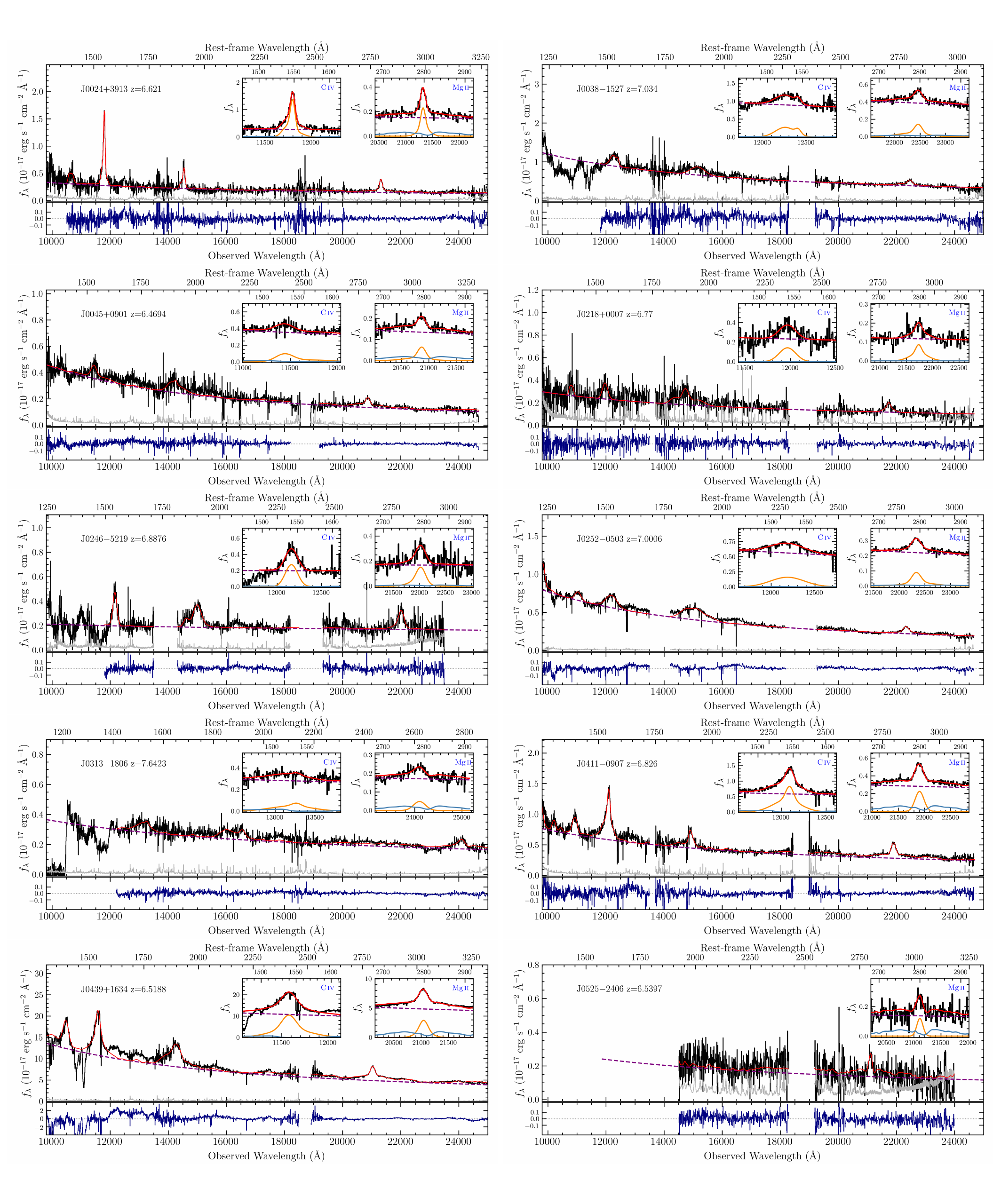} 
\caption{Model fits to the spectra of quasars in this sample, other than J0139--1008. The black and grey lines are the observed spectrum and spectral uncertainty. The purple dashed line represents the best fit power law continuum and the solid red line denotes the total fit. The two inset plots show the fits to the \civ\ and \mgii\ emission lines. The orange and blue solid lines represent the best-fit emission line and iron component, respectively. The redshifts shown here are from [\cii] (where available) or from \mgii\ (for objects without [\cii] observations). The bottom panel (dark blue line) shows the residual (data -- model) of the spectral fitting.}
\label{fig:allspec01}
\end{figure*}

\addtocounter{figure}{-1}
\begin{figure*}%
\centering 
\epsscale{1.2}
\plotone{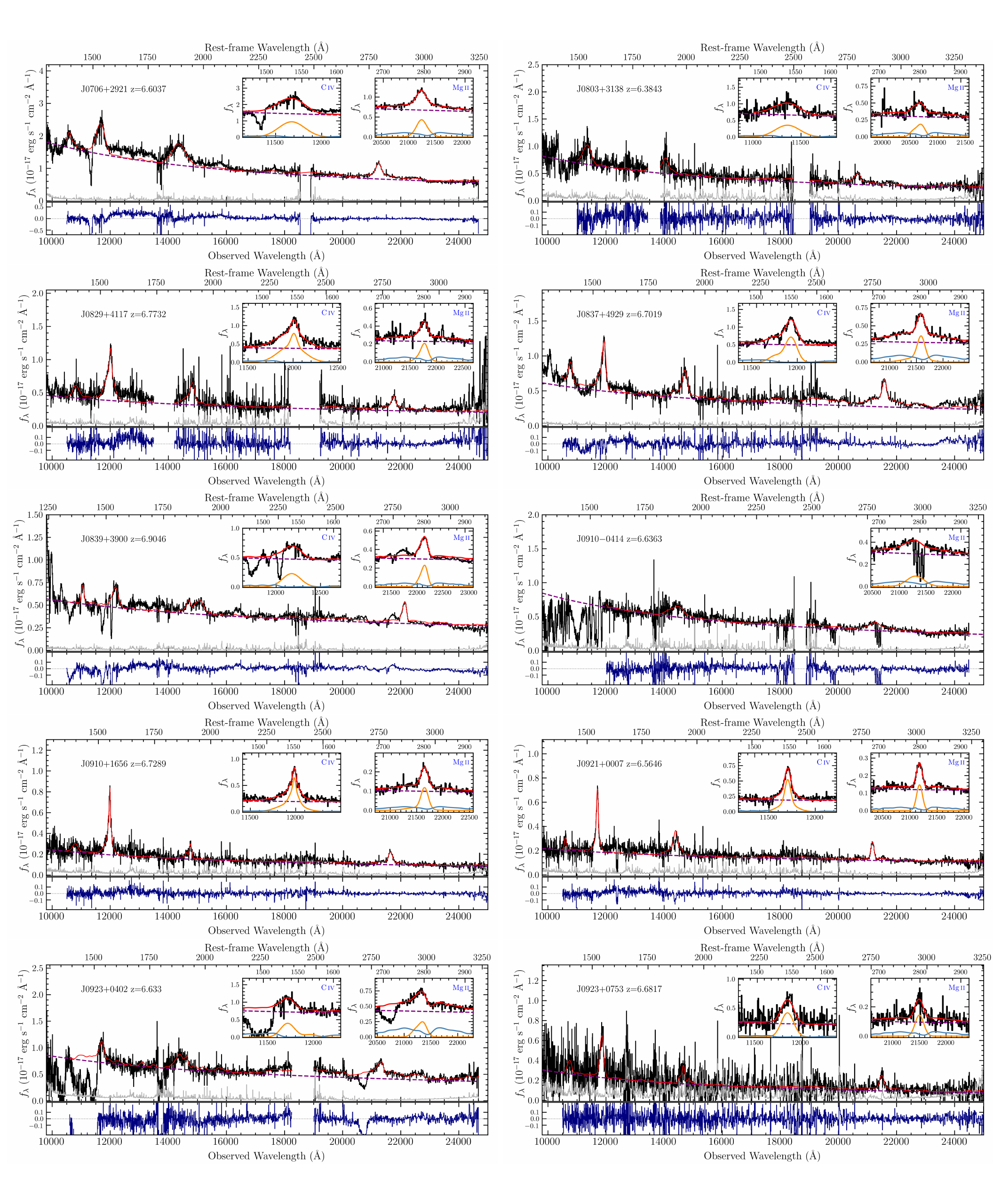} 
\caption{Continued.}
\label{fig:allspec02}
\end{figure*}

\addtocounter{figure}{-1}
\begin{figure*}%
\centering 
\epsscale{1.2}
\plotone{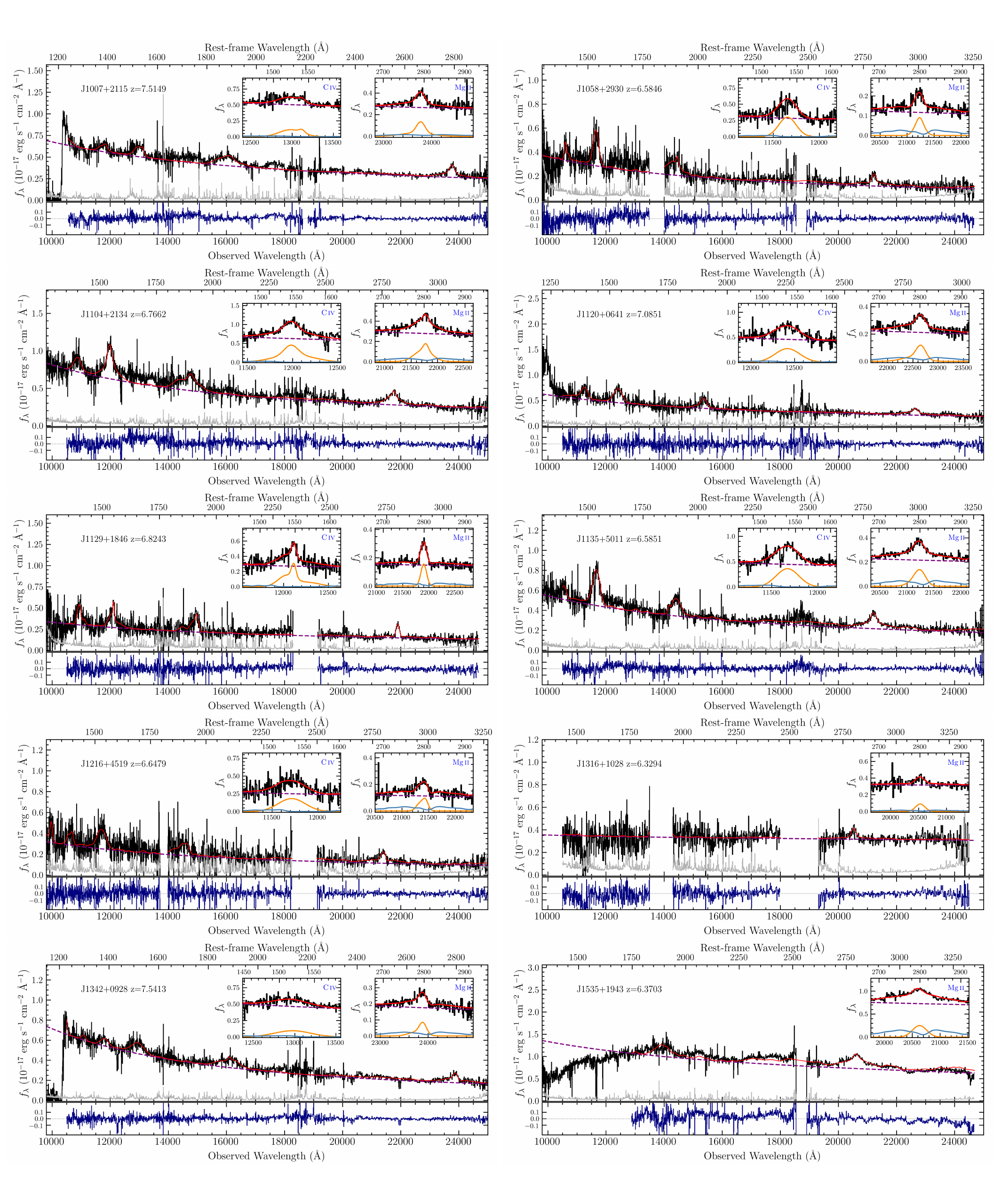} 
\caption{Continued.}
\label{fig:allspec03}
\end{figure*}

\addtocounter{figure}{-1}
\begin{figure*}%
\centering 
\epsscale{1.2}
\plotone{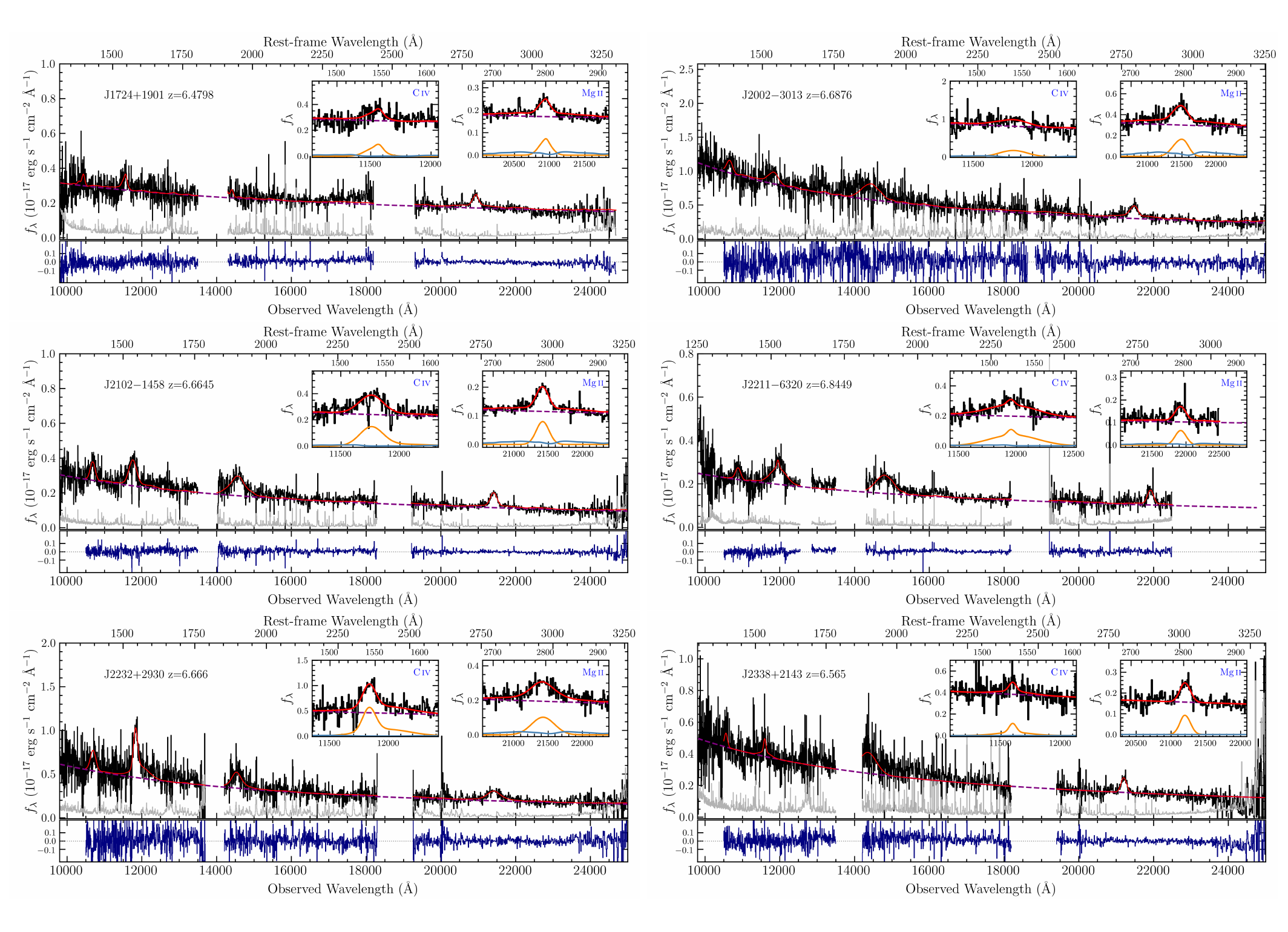} 
\caption{Continued.}
\label{fig:allspec04}
\end{figure*}

\section{Measurements Using the VW01 Iron Template}
We present the BH masses measured based on the spectral fitting using the VW01 iron template, as a comparison.

\begin{deluxetable}{ l l l l l}
\tablecaption{Quasar BH masses derived from the spectral fitting based on the VW01 iron template.}\label{tab:vw01fitting}
\tabletypesize{\scriptsize}
\tablewidth{0pt}
\tablehead{
\colhead{Name} &
\colhead{$L_{\rm Bol}$} &
\colhead{FWHM$_{\rm MgII}$} &
\colhead{$M_{\rm BH}$} &
\colhead{$\lambda_{\rm Edd}$}
\\
\colhead{}  &  \colhead{($10^{46} $\rm erg s$^{-1}$)} & \colhead{(km s$^{-1}$)} & \colhead{($10^{9}\,M_{\odot}$)} & \colhead{}
}
\startdata
J0024+3913  &  8.7$\pm$0.8  &  1783$\pm$38  &  0.299$\pm$0.001  &  2.3$\pm$0.2 \\
J0038--1527  &  23.8$\pm$1.0  &  3102$\pm$45  &  1.50$\pm$0.08  &  1.3$\pm$0.1 \\
J0045+0901  &  6.7$\pm$0.6  &  2911$\pm$131  &  0.70$\pm$0.03  &  0.8$\pm$0.1 \\
J0218+0007  &  6.4$\pm$1.4  &  2745$\pm$42  &  0.61$\pm$0.07  &  0.8$\pm$0.2 \\
J0246--5219  &  10.2$\pm$1.0  &  3319$\pm$693  &  1.12$\pm$0.50  &  0.7$\pm$0.3 \\
J0252--0503  &  13.2$\pm$0.4  &  3406$\pm$219  &  1.34$\pm$0.20  &  0.8$\pm$0.1 \\
J0313--1806  &  14.2$\pm$0.7  &  4219$\pm$465  &  2.14$\pm$0.52  &  0.5$\pm$0.1 \\
J0319--1008  &  9.7$\pm$1.4  &  2103$\pm$8  &  0.44$\pm$0.04  &  1.8$\pm$0.3 \\
J0411--0907  &  16.0$\pm$1.1  &  2837$\pm$75  &  1.03$\pm$0.09  &  1.2$\pm$0.1 \\
J0439+1634  &  5.1$\pm$0.1  &  3041$\pm$14  &  0.668$\pm$0.003  &  0.6$\pm$0.1 \\
J0525--2406  &  7.7$\pm$3.9  &  2048$\pm$472  &  0.37$\pm$0.07  &  1.6$\pm$0.9 \\
J0706+2921  &  36.8$\pm$1.6  &  3375$\pm$31  &  2.20$\pm$0.01  &  1.3$\pm$0.1 \\
J0803+3138  &  14.5$\pm$1.3  &  4432$\pm$119  &  2.39$\pm$0.07  &  0.5$\pm$0.1 \\
J0829+4117  &  13.0$\pm$1.0  &  2869$\pm$82  &  0.95$\pm$0.04  &  1.1$\pm$0.1 \\
J0837+4929  &  17.4$\pm$0.5  &  2565$\pm$48  &  0.88$\pm$0.02  &  1.6$\pm$0.1 \\
J0839+3900  &  18.7$\pm$0.6  &  2332$\pm$47  &  0.75$\pm$0.02  &  2.0$\pm$0.1 \\
J0910--0414  &  15.9$\pm$0.9  &  7825$\pm$844  &  7.81$\pm$1.45  &  0.2$\pm$0.1 \\
J0910+1656  &  5.9$\pm$0.6  &  2321$\pm$40  &  0.42$\pm$0.04  &  1.1$\pm$0.1 \\
J0921+0007  &  6.6$\pm$0.6  &  1729$\pm$105  &  0.24$\pm$0.04  &  2.1$\pm$0.4 \\
J0923+0402  &  25.7$\pm$3.1  &  3362$\pm$183  &  1.83$\pm$0.21  &  1.1$\pm$0.2 \\
J0923+0753  &  5.7$\pm$1.8  &  2800$\pm$475  &  0.60$\pm$0.10  &  0.8$\pm$0.3 \\
J1007+2115  &  20.4$\pm$1.4  &  3321$\pm$170  &  1.59$\pm$0.17  &  1.0$\pm$0.1 \\
J1058+2930  &  6.5$\pm$1.5  &  2642$\pm$76  &  0.57$\pm$0.10  &  0.9$\pm$0.3 \\
J1104+2134  &  15.8$\pm$0.9  &  4198$\pm$55  &  2.24$\pm$0.01  &  0.6$\pm$0.1 \\
J1120+0641  &  13.4$\pm$1.1  &  3928$\pm$344  &  1.80$\pm$0.28  &  0.6$\pm$0.1 \\
J1129+1846  &  9.1$\pm$2.0  &  1862$\pm$45  &  0.33$\pm$0.02  &  2.2$\pm$0.5 \\
J1135+5011  &  12.1$\pm$0.8  &  3651$\pm$98  &  1.48$\pm$0.03  &  0.6$\pm$0.1 \\
J1216+4519  &  6.5$\pm$1.2  &  3102$\pm$652  &  0.78$\pm$0.42  &  0.7$\pm$0.4 \\
J1316+1028  &  15.2$\pm$2.9  &  3168$\pm$1314  &  1.25$\pm$0.80  &  1.0$\pm$0.6 \\
J1342+0928  &  13.3$\pm$1.1  &  3094$\pm$195  &  1.12$\pm$0.15  &  0.9$\pm$0.2 \\
J1535+1943  &  36.9$\pm$1.8  &  6577$\pm$464  &  8.39$\pm$0.98  &  0.3$\pm$0.1 \\
J1724+1901  &  8.6$\pm$1.3  &  2927$\pm$50  &  0.80$\pm$0.07  &  0.9$\pm$0.1 \\
J2002--3013  &  16.5$\pm$1.9  &  3501$\pm$250  &  1.59$\pm$0.21  &  0.8$\pm$0.1 \\
J2102--1458  &  6.3$\pm$0.6  &  2882$\pm$203  &  0.66$\pm$0.13  &  0.7$\pm$0.2 \\
J2211--6320  &  5.8$\pm$0.2  &  2613$\pm$819  &  0.53$\pm$0.36  &  0.9$\pm$0.6 \\
J2232+2930  &  10.4$\pm$1.6  &  6705$\pm$810  &  4.63$\pm$1.13  &  0.2$\pm$0.1 \\
J2338+2143  &  7.6$\pm$1.3  &  2434$\pm$113  &  0.52$\pm$0.03  &  1.2$\pm$0.2 \\
\enddata
\end{deluxetable}

\end{document}